\documentclass{emulateapj}
\usepackage{apjfonts}
\usepackage{psfig}
\def\gs{\mathrel{\raise0.35ex\hbox{$\scriptstyle >$}\kern-0.6em 
\lower0.40ex\hbox{{$\scriptstyle \sim$}}}}
\def\ls{\mathrel{\raise0.35ex\hbox{$\scriptstyle <$}\kern-0.6em 
\lower0.40ex\hbox{{$\scriptstyle \sim$}}}}
\def\et{\hbox{et al.}$\,$}
\def\etal{\hbox{et al.}}
\def\OII{\hbox{[O II]}$\,\,$}

\def\Ha{\hbox{H$\alpha$}$\,$}

\def\Msun{\rm{\hbox{M$_{\odot}$}}}           
\def\ang{\hbox{$\,$\AA}}
\def\kms{\rm{\hbox{km s$^{-1}$}}}
\def\OII{\hbox{[O II]}$\,$}
\def\OIII{\hbox{[O III]}$\,$}

\def\24m{\hbox{24~$\micron$}$\,$}
\def\10-18{\hbox{$\times~10^{-18}$}}
\def\deg{\hbox{$^{\circ}$}}

\def\lya{\mbox {Ly$\alpha$~}}
\def\Lya{\mbox {Ly$\alpha$}}
\def\flux{~ergs~s$^{-1}$~cm$^{-2}$}
\def\lum{~ergs~s$^{-1}$}
\def\kms{~km~s$^{-1}$~}

\slugcomment{Submitted: 2011 March 24; accepted: 2011 July 20}

\shortauthors{Dressler \etal\ }
\shorttitle{Faint \Lya\ Emitters}

\begin{document}

\title{Detections of Faint \Lya\ Emitters AT \lowercase{z} = 5.7: Galaxy Building Blocks and Engines of Reionization\footnote{T\lowercase{his 
paper includes data gathered with the 6.5 meter \uppercase{M}agellan \uppercase{T}elescopes located at \uppercase{L}as 
\uppercase{C}ampanas \uppercase{O}bservatory, \uppercase{C}hile.}}}

\author{Alan Dressler}
\affil{Carnegie Observatories, 813 Santa Barbara St., Pasadena, California 91101-1292}
\email{dressler@obs.carnegiescience.edu}

\author{Crystal L. Martin and Alaina Henry}
\affil{University of California, Santa Barbara, Department of Physics, Santa Barbara, CA 93106}
\email{cmartin@physics.ucsb.edu; ahenry@physics.ucsb.edu}

\author{Marcin Sawicki}
\affil{St. Mary's University, Department of Astronomy and Physics, 923 Robie Street,
Halifax, N.S., B3H 3C3, Canada}
\email{sawicki@ap.smu.ca}

\author{Patrick McCarthy}
\affil{Carnegie Observatories, 813 Santa Barbara Street, Pasadena, California 91101-1292}
\email{pmc2@obs.carnegiescience.edu}

\begin{abstract}

We report results of a unprecedentedly deep, blind search for \Lya\ emitters (LAEs) at $z = 5.7$ using \emph{IMACS}, 
the \emph{Inamori-Magellan Areal Camera \& Spectrograph}, with the goal of identifying missing sources 
of reionization that could also be basic building blocks for today's $L^*$ galaxies.  We describe how improvements in wide 
field imaging with the Baade telescope, upgrades to \emph{IMACS}, and the accumulation of $\sim$20 hours of integration 
per field in excellent seeing led to the detection of single-emission-line sources as faint as $F \approx 2\10-18$ \flux, a 
sensitivity 5 times deeper than our first search (Martin \etal\ 2008).  A reasonable correction for foreground interlopers 
implies  a steep rise of approximately an order of magnitude in source density for a factor of four drop in flux, from 
$F = 10^{-17.0}$\flux\ to $F = 10^{-17.6}$ ($2.5\, \10-18$)\flux.  At this flux the putative LAEs have reached 
a surface density of $\sim$1 per sq arcminute --- a comoving volume density of $4\,\times10^{-3}$ Mpc$^{-3}$, several times 
the density of \emph{L*} galaxies today.   Such a population of faint LAEs would account for a significant fraction of the 
critical flux density required to complete reionization at this epoch, and would be good candidates for building blocks of 
stellar mass $\sim10^{8-9}$ \Msun\ for the young galaxies of this epoch.


\end{abstract}



\section{Introduction}

It is known, from the observed polarization of the cosmic microwave background (Dunkley \etal\ 2009; Komatsu 
\etal\ 2009), that reionization of neutral hydrogen in the intergalactic medium (IGM) started at $z\gs8$ and, 
from the spectra of high-redshift quasars (Fan \etal\ 2002), was largely complete by $z\sim6$.  However,  the 
formulation of a detailed picture describing, for example, when reionization began and how rapidly it 
proceeded, remains in an early stage.  Even the source of the ionizing photons has not been 
\emph{definitively} identified, although star-forming galaxies are the favored candidates. 

Efforts to demonstrate whether galaxies can produce sufficient Lyman continuum radiation to maintain an ionized 
IGM began with the discovery of galaxies at $z\sim6$ (Bouwens \etal\ 2004; Bunker \etal\ 2004; Dickinson 
\etal\ 2004; Yan \& Windhorst 2004)  and the  identification of this redshift as the epoch when reionization was 
completed (from Gunn-Peterson troughs in quasar spectra --- see Fan \etal\ 2002).   It was quickly recognized 
that the observed galaxies do not produce enough ionizing photons at $z\sim6$ to balance the recombination rate 
in the IGM.  However, this interpretation is complicated by the difficulty of predicting the ionizing output of galaxies 
from their 1500\ang\ UV-continuum (Steidel \etal\ 2001; Shapley \etal\ 2006; Malkan \etal\ 2003; Siana \etal\ 2007, 
2010; Cowie \etal\ 2009; Iwata \etal\ 2009; Vanzella \etal\ 2010; Nestor \etal\ 2011) and of estimating the clumpiness 
of the IGM (Madau \etal\ 1999; Pawlik \etal\ 2009) at these early epochs.  Regardless, galaxies that are fainter 
than the limits of current surveys surely exist, and could in fact be the dominant source of ionizing photons 
(Bunker \etal\ 2004; Yan \& Windhorst 2004; Trenti \etal\ 2010;  Salvaterra \etal\ 2010). 

Quantifying faint populations that remain undetected is crucial if we are to determine whether star-forming galaxies 
reionized the IGM.  The most efficient way to find the faintest galaxies is to search for their strong \Lya\ emission, since such 
galaxies can be detected even when their stellar continuum radiation is fainter than the detection limit of the deepest 
\emph{HST} surveys.  Cowie \& Hu (1998) pioneered the effort of using a  narrowband filter to find $z\sim3$ \Lya\ emitting 
galaxies (LAEs). More than a decade later newer surveys have succeeded at higher redshifts and taken advantage of
large telescopes with larger fields of view, notably SuprimeCam on the Subaru telescope (Miyazaki \etal\ 2002).   Large 
samples have been compiled for LAEs at $z=5.7$ and $6.5$, where \Lya\ emission is redshifted into `windows' of 
the OH airglow spectrum centered at $\lambda \approx 8200$\ang\ and $\lambda \approx 9200$\ang\ (Malhotra \& 
Rhoads 2002, 2004; Shimasaku \etal\ 2006; Kashikawa \etal 2006, 2011; Hu \etal\ 2004, 2010; Ouchi \etal\ 2008, 2010).  
Nevertheless, only a fraction of these narrowband imaging selected sources have been spectroscopically confirmed. 

The contribution of LAEs to the ionizing photon budget has been debated.  At $z\sim3$ only 25\% of star-forming 
Lyman Break Galaxies (LBGs) have \Lya\ emission with an equivalent width large enough to be included in narrowband 
samples (Shapley \etal\ 2003; cf, Steidel \etal\ 2011).    However, the prevalence of \Lya\ emission in LBGs samples 
appears to grow with increasing redshift.  This was initially implied by the different evolutionary paths of the \Lya\ luminosity 
function of LAEs (Ouchi \etal\ 2008)  and the UV luminosity function of LBGs (Bouwens \etal\ 2007).  Spectroscopic followup 
of $z\sim4-6$ LBGs by Stark \etal\ (2010, 2011) seems to support this interpretation, but Douglas \etal\ (2010) reach
a different conclusion.  Regardless, if the Lyman-continuum emission is higher in LAEs than their LBG counterparts 
(as suggested by Iwata \etal\ 2009, Inoue \etal\ 2011, and Nestor \etal\ 2011) then observations could ultimately prove 
that LAEs dominate the reionization. Therefore, it is especially important to quantify the faint-end of the \Lya\ luminosity 
function back into the epoch of reionization.  

Finding faint LAEs at $z\sim6$ requires a new approach to reach fainter than is possible with narrowband imaging.  
Spectroscopic searches accomplish this by dispersing the sky spectrum and observing  at a spectral resolution 
that is close to the width of the emission line.   emission-line searches have now been carried out using a long-slit 
(Rauch \etal\ 2008), serendipitously (Sawicki \etal\ 2008; Lemaux \etal\ 2009; Cassata \etal\ 2011), and multi-slit 
``venetian blind'' masks (Martin \& Sawicki 2004; Tran \etal\ 2004).  In Martin \etal\ (2008, hereinafter, Martin08) 
we presented the first sample of three $z\sim6$ LAEs to be found by this multi-slit method and confirmed as LAEs 
in followup spectroscopy.  These observations reached depths comparable to the narrowband imaging surveys at 
this redshift (Ouchi \etal\ 2008; Murayama \etal\ 2007; Shimasaku \etal\ 2006; Hu \etal\ 2010), and they demonstrated 
that a fainter survey could be carried out.  

We have now completed a deeper search for \Lya\ emitters at $z=5.7$ and achieved a sensitivity about 5 times 
better than our previous survey (Martin08).  Among our candidates are many low-luminosity LAEs --- 
only a few such objects have been discovered previously, via the technique of strong lensing by rich clusters of 
galaxies (Santos \etal\ 2004).  We present the initial results of this search, principally a steep rise in 
the number of LAE candidates to faint fluxes. In \S 2 we describe improvements in our observational search 
with \emph{IMACS} that led to marked improvement compared to our earlier efforts.  In \S 3 we describe the data 
used in this study,  and in \S 4, we derive source counts for detected objects that show a steep rise in faint 
single-line sources.  In \S5 we present evidence, based on angular correlation functions and subtraction of 
known foreground populations, that a significant fraction of these faint sources are in fact LAEs.  In \S6 we discuss 
the implications of this result for the issues of early generations of star formation, the building blocks of the first 
galaxies, and the sources of cosmic reionization.

We adopt cosmological parameters of $\Omega_m = 0.30$, $\Omega_{\Lambda} = 0.7$, and $H_{0}  = 70$\kms 
Mpc$^{-1}$ throughout.  

\section{Improvements in our NBMS measurements}

The technique of `multislit narrow-band spectroscopy' (MNS) was pioneered by Crampton \& Lilly (1999) and
Martin \& Sawicki (2004).  We developed the technique for \emph{IMACS} starting in 2004, conducting a search in both
2004 and 2005 in two fields, a 10h field in area of the COSMOS Survey (Scoville \etal\ 2007), and the other a 15h 
field from the Las Campanas Infrared Survey (Marzke \etal\ 1999).  We used a ``venetian-blind'' mask with 1.5\arcsec\
wide slits that covered the full $\approx$28\arcmin\ diameter field of the \emph{IMACS} f/2 camera with a filling 
factor of 10\%. In that first application of MNS, we accumulated 5-10 hours of integration at two positions in each 
of these two fields, in photometric conditions that produced a point-source image size of $\sim$0.8\arcsec\ FWHM.  
Further details of the observational setup are given in Martin08.

Our motivation for using the MNS technique was to carry out a deeper \Lya\ search than had been 
reached with narrow-band imaging, $F = 10^{-17}$\flux, the limiting flux that was reached in the
Subaru Deep Field (SDF, Shimasaku \etal\ 2006), corresponding to a magnitude in the narrow-band 
filter NB816 filter of $NB816_{AB} = 26.0$.  The depth advantage of MNS can be understood 
by comparing how sky background affects the limiting flux of source detection.  With our MNS technique, 
a source is superposed on the dispersed sky background, which for typical \Lya\ emitters is 10-15\ang\ of spectrum.
In comparison, a narrow-band detection must compete against sky background approximately ten times
greater.  Using the 2008 MNS exposures discussed here, we found that, for a source $F = 10^{-17}$\flux, 
summed over 2\arcsec\ in the spatial direction and the $\sim120$\ang\ FWHM bandpass of the NB816 filter, the ratio of 
counts (detected photoelectrons) $N_{sky}/N_{obj} \approx 100$.  Thus, going fainter than this limit requires exceeding 
the precision of 1\% photometry.  Stubbs and Tonry (2006) describe in detail why, for standard \emph{ground-based} 
observations with CCD detectors, achieving better than 1\% photometry precision is not possible in practice.\footnote{Systematic 
errors of ground-based observations --- chief among them, flat fielding errors due to differences in the energy distributions 
of `flats', (time-dependent) sky, and sources --- limits precision to 1\% at best unless `chopping' techniques are employed.}
The hardness of this limit can be appreciated by inspecting Figure 1 of Takahashi \etal\ (2007),  which shows the rapid 
rise in photometric errors as $NB816_{AB} = 26.0$ is approached.  By reducing the sky background to what is effectively
a narrower bandpass of 10-15\ang, our MNS observations with \emph{IMACS} reach a line flux an order-of-magnitude 
deeper before this photometric limit is reached. 

Our initial attempt successfully demonstrated MNS with \emph{IMACS} and was the first time high-redshift \Lya\ 
emitters had actually been detected with the technique.  However, our first survey had not reached any 
fainter than the limit of narrow-band imaging surveys, which have the considerable advantage of covering
large areas efficiently.  We learned that a number of performance issues were responsible for limiting
the sensitivity of our first search well below the potential of the technique.  During the next two 
years, as we made follow-up observations for the three \Lya\ detections and other candidates presented
in Martin08, we worked to improve several aspects of the system in an attempt to reach the project's original 
goal of $F \approx 3 \times 10^{-18}$\flux.  Because of the unusual gain in sensitivity of a factor of $\sim$5 
between our earlier survey and the 2008 survey reported here, we briefly describe these improvements:
 
\begin{itemize}

\item{An undetected misalignment of Magellan-Baade wide-field ADC (atmospheric-dispersion-compensating)
corrector delivered a significantly tilted focal plane to \emph{IMACS} until 2007. This resulted in aberrated 
images for \emph{IMACS} and also prevented the active optics system, which relies on the \emph{IMACS} 
guiders, to provide accurate information  to correct primary mirror shape, telescope collimation and focus, and 
to accurately control telescope tracking.  Once the ADC-corrector was properly aligned, the performance of 
the system improved dramatically.}

\item{The ``as-built'' optics of the \emph{IMACS} f/2 camera limited image quality to about 0.7" FWHM over the 1/2-deg 
field, compared to the specification of a 0.35\arcsec\ FWHM contribution to the image point spread function (\emph{psf}).  
Tests performed in 2007 showed that a field-dependent coma was mainly responsible, with additional degradation from focal 
plane tilt and astigmatism.  Optical modeling led to a repositioning of the field flattener in order to cancel the coma and reduce 
astigmatism (see Dressler \etal\ 2011 for a full description).  Though a difficult modification at that stage of the \emph{IMACS} 
operation, the resulting adjustments returned the camera to performance close to original specifications: in the best seeing, 
({\it psf} $\le$0.35\arcsec\ FWHM), the f/2 camera  produces 0.50\arcsec\ FWHM images over 85\% of field.}  

\item{With the substantial help of a 2006 NSF {\it TSIP} award, the original CCD Mosaic camera at the f/2
focus was replaced by a new one using E2V detectors.  This raised the system throughput (including 
telescope) from 14\% to 22\% at 8200\ang, a factor of 1.6 gain.}

\item{The first 4 nights of a 5-night observing run with \emph{IMACS} for this program, April 8-12, 2008, were 
excellent --- clear, with seeing of typically 0.45 - 0.55-arcsec {\it on the detector}, and with perfect performance of 
telescope and instrument.  The quality of the data collected on these nights far surpassed that of the 2004-5 data.}

\end{itemize}

\section{Data and Data Reduction}

Our observational setup in 2008 followed that of our earlier \Lya\ search, described in Martin08.  One new slitmask
was fabricated for the 10h field; the new search mask copied the original design of 100 parallel slits of width 1.5\arcsec\ 
and a center-to-center spacing of 15\arcsec.  As before, this layout sampled 10\% of the {IMACS f/2 field}.  A custom 
narrow-band filter centered at 8190\ang\ with a designed FWHM $\Delta\lambda \approx 150$\ang\ was mounted in the 
\emph{IMACS}\, pupil directly in front of the 200-l/mm grism.  This configuration produces 100 spectra, at a dispersion 
of 2.0\ang\ pix$^{-1}$, and a projected 1.5\arcsec\ slit-width of 6 pixels. The spectra overlap slightly in wavelength, and 
there is a zero-order image of the slitmask (and field) that covers much of the lower half (chips 1-4) of the CCD mosaic, 
complicating the reduction.  A small amount of sky coverage is also lost because of metal bridges (`tines') that stiffen the 
mask against warping by  dividing the long slits into discrete sections.   After accounting for these obstructions, the actual 
on-sky area is 55.3 squarearcminutes per exposure.  A schematic of the search mask is shown in Figure 2 of Martin08.

Over the nights April 8-12, 2008, 24 frames were taken in the 10h COSMOS field (10:00:43, 02:11:00 [2000]) and
23 in the 15h LCRIS (15:23:35, -00:08:00 [2000]) field, for a total exposure of 18.75 and 22.75 hours, respectively.  
For each night's exposure there was a slight drift of the slitmask pattern over the detector, about 5 pixels along the 
dispersion and 2 pixels across, that accumulated as the field was tracked from several hours east to several hours 
west of the meridian.  After dome-flats were used to correct the pixel-to-pixel sensitivity variations of the detectors, 
these frames were shifted before stacking to form a single, very deep exposure for each of the two fields.  Because 
this was a blind search, object positions were not known {\it a priori}, so the \emph{IMACS} data reduction package 
{\it COSMOS}\footnote{http://www.obs.carnegiescience.edu/Code/cosmos} could not be used at this stage.  Instead, a set 
of {\it IRAF} scripts were developed, the most important of which was a ``running-average'' of $\pm10$ columns
(the dispersion direction) that provided a mean sky spectrum to subtract sky from each of the 100 long-slit spectral ``bands'' 
(short in wavelength, long on the sky).  This worked well, although it was recognized that more accurate sky subtraction 
would need to be redone later ---after object identification --- because the sky determined with the running-average 
technique is necessarily biased by as-yet-unidentified objects that are included in the averages.

Because of the high number of artifacts and non-astronomical ``features'' on these unconventional data frames, 
we chose to search by eye for emission-line objects, with and without continuum to the blue of the emission.  Martin, 
Dressler, and McCarthy independently examined both fields, compiling lists that separated the detections into four 
categories: (1) clear single emission-line source without a blue continuum; (2) probable but not certain single 
emission-line source without a blue continuum; (3) single or multiple emission-lines source with a blue 
continuum\footnote{A few multiple emission-line sources without blue continua were included in this category.}; 
(4) possible, low S/N detection of an emission line without a blue continuum. The last category had the largest 
source count and probably contained comparable numbers of real and spurious detections.

The sample used in this paper includes only those objects that are certainly real or very likely to be; almost all were originally 
placed by the three classifiers in classes 1 \& 2.  After cross-comparing and re-examining our individual lists, a final list of such 
candidates was assembled.  The 10h field has 104 single-emission-line sources without blue continua (80 class (1) 
sources and 24 class (2) sources) and 130 emission-line sources with blue continua. The 15h field has 111 single-emission-line 
sources without blue continua (77 class (1) sources and 34 class (2) sources) and 105 emission-line sources with blue 
continua. We will call the latter `Em+C' --- sources in this category are objects that are certainly foreground (including the few 
cases of multiple emission-line sources without a detectable continuum.)  However, the single-emission-line sources without 
blue continua, which we will call SELs, are candidate LAEs. Figure 1 shows a mosaic of $5\arcsec \times 50\ang$ 
spectra for the SELs  in the 10h COSMOS field, and Figure 2 shows the SELs in the 15h LCRIS field.

In Figure 3 we show the derived signal-to-noise ratio (S/N) for the sources shown in Figures 1 \& 2.  These were determined
by adding up the flux in square apertures --- 1.8\arcsec\ for sources brighter than, and 1.4\arcsec\ for sources fainter than, 
$F = 10^{-17.5}$ ($3.2\,\10-18$)\flux --- and by estimating the noise from the much greater sky flux, determined separately for 
each of the two fields.  (The choice of slightly smaller apertures for the fainter sources modestly reduces the scatter in S/N ratios.)
The plot confirms that incompleteness 

\begin{figure*}
\vspace*{0.1cm}
\hbox{~}

\centerline{\psfig{file=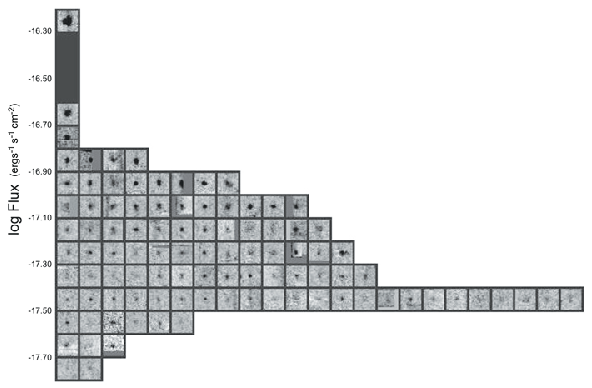,angle=0.0,width=6.0in}}
\noindent{\scriptsize
\addtolength{\baselineskip}{-3pt}
 
\hspace*{0.3cm}Fig.~1.\  Mosaic of 5\arcsec\ $\times$ 50\ang\ boxes (spatial -- horizontal, wavelength -- vertical) of the sample
of SELs (single-emission-line without blue continuum) sources in the 10h (COSMOS) field.   A flux of $F = 10^{-17}$\flux, a natural limit for narrow-band photometric studies, is easily reached with these data, at the $\sim10\sigma$ level.  A flux of 
$F = 3\,\10-18$\flux\ is detected at the $\sim3-4\sigma$ level.  The number of sources rises rapidly with decreasing flux, 
as described in the text.  Strong artifacts (eg., cosmic rays, bad pixels, and sky subtraction problems) have been blanked
out in some cases. 

\vspace*{0.2cm}
\addtolength{\baselineskip}{3pt}
}

\end{figure*}


\begin{figure*}
\vspace*{0.1cm}
\hbox{~}

\centerline{\psfig{file=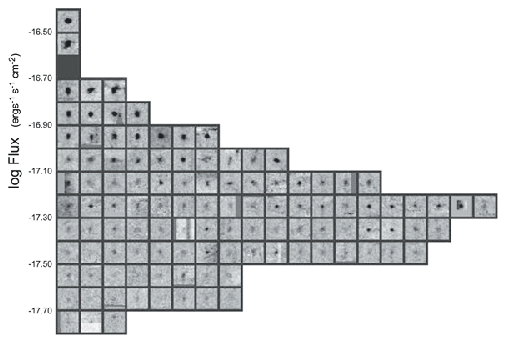,angle=0.0,width=6.0in}}
\noindent{\scriptsize
\addtolength{\baselineskip}{-3pt}
 
\hspace*{0.3cm}Fig.~2.\  Same as Figure 1, for detected SEL sources in the 15h LCRIS field.

\vspace*{0.2cm}
\addtolength{\baselineskip}{3pt}
}

\end{figure*}

\noindent{becomes substantial at log $F$ ergs s$^{-1}$ cm$^{-2} \approx -17.6$, ($2.5\,\10-18$\flux), at 
which point the typical $S/N \approx 3$.  Although our selection was done visually, we were apparently 
successful in drawing the line with a relatively sharp cutoff in S/N.  The implication, of course, is that there 
are many more real sources among the class 4 candidates, but also many that are statistical fluctuations 
and artifacts, as we supposed.  In this paper we use a flux cut of log $F$ (ergs s$^{-1}$ cm$^{-2}$) $\ge -17.6$.


\vspace*{0.1cm}
\hbox{~}

\centerline{\psfig{file=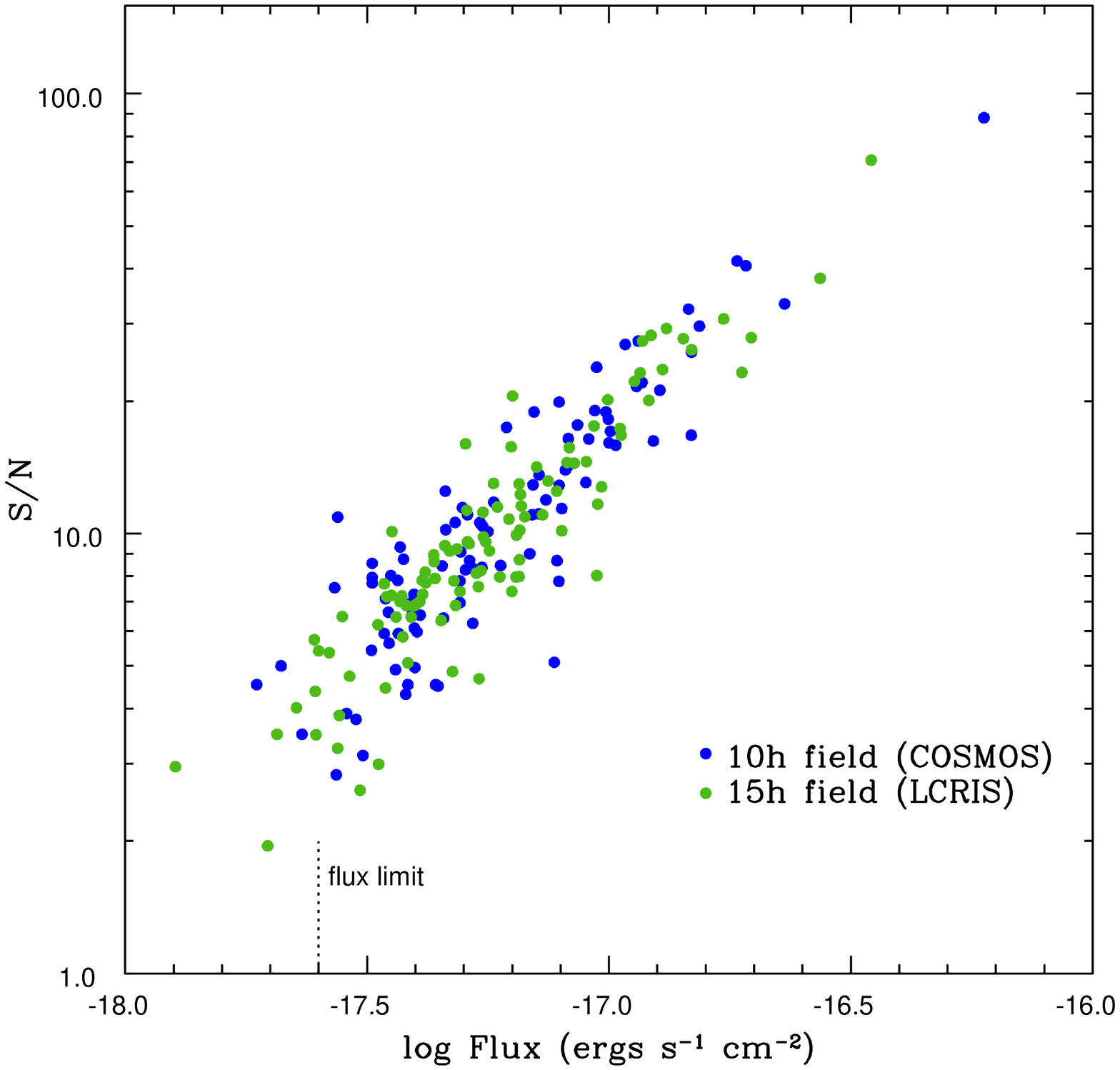,angle=0.0,width=3.2in}}
\noindent{\scriptsize
\addtolength{\baselineskip}{-3pt}
 
\hspace*{0.3cm}Fig.~3.\  S/N ratios for the detections displayed in Figures 1 \& 2.  blue points -- 10h field; 
green points -- 15h field.  The sample used in this paper is cut at log $F$ (ergs s$^{-1}$ cm$^{-2}$) $> -17.6$ ($2.5\,\10-18$\flux).

\vspace*{0.2cm}
\addtolength{\baselineskip}{3pt}
}


Once the SEL and Em+C sources were identified, the {\it COSMOS} data reduction package could be used to make
2-D spectral extractions with accurate sky subtraction.  Another {\it COSMOS} program, {\it viewspectra}, was used
to examine the sky-subtracted 2-D spectrum in order to select the spectral ``columns'' containing the object.
Generally, the extractions were 9 pixels centered on the object (1.8\arcsec) but some variation was made to
optimize signal-to-noise ratio for more diffuse or more compact objects, and to avoid artifacts that dot the frames.
These 1-D spectra were then inspected with {\it IRAF} {\it splot} to make, if necessary, an adjustment to the
continuum level, in order to measure the emission-line over zero-backround, and to determine the wavelength limits 
for integrating the counts.  Counts were summed and converted to flux by the calibration $1.35 \times 10^{-20}$\flux\ 
per \emph{IMACS} count (photoelectron).  The calibration came from 6 measurements of 2 Hamuy standard stars 
EG274 and LTT7379 (Hamuy \etal\ 1994) through a 7\arcsec\ round aperture on two of the four nights, which were 
in agreement to about 5\%, a systematic error considerably larger than the photon statistics of the observations.

\section{Results of the Search: Derivation of Source Counts}

With measured fluxes for the SEL sources --- presumed $z\sim5.7$ \Lya\ emitters (LAEs) or foreground 
emission-line galaxies, and fluxes for the Em+C sources --- certain foreground galaxies, we plot in Figure 4 
cumulative source counts versus measured flux.  The blue line shows the cumulative distribution of all 
emission-line sources detected in the two fields.  The black line shows the counts for SEL sources.  The 
black line rises steeply, arguably more steeply than any known foreground population, as indicated, for example by 
our own counts of foreground galaxies, the Em+C sources --- the red line.  (As we show in the next section, these 
are almost entirely \OII\ emitters at $z \approx 1.19$, \OIII\ emitters at $z = 0.63$, and \Ha\ at $z = 0.25$.)  This is 
our first indication that many of the SEL sources are LAEs at $z = 5.7$.  However, another possibility is that 
foreground galaxies that would otherwise be on the red line have been added to the black line, if we systematically 
fail to detect their continuum flux as we observe fainter sources of moderate equivalent width. For this reason, 
we cannot simply credit the steep rise in emission-line-only sources as exclusively or even \emph{mainly} due 
to LAEs.  We return to this issue below after discussing the incompleteness correction for the SEL counts.


\vspace*{0.1cm}
\hbox{~}
\centerline{\psfig{file=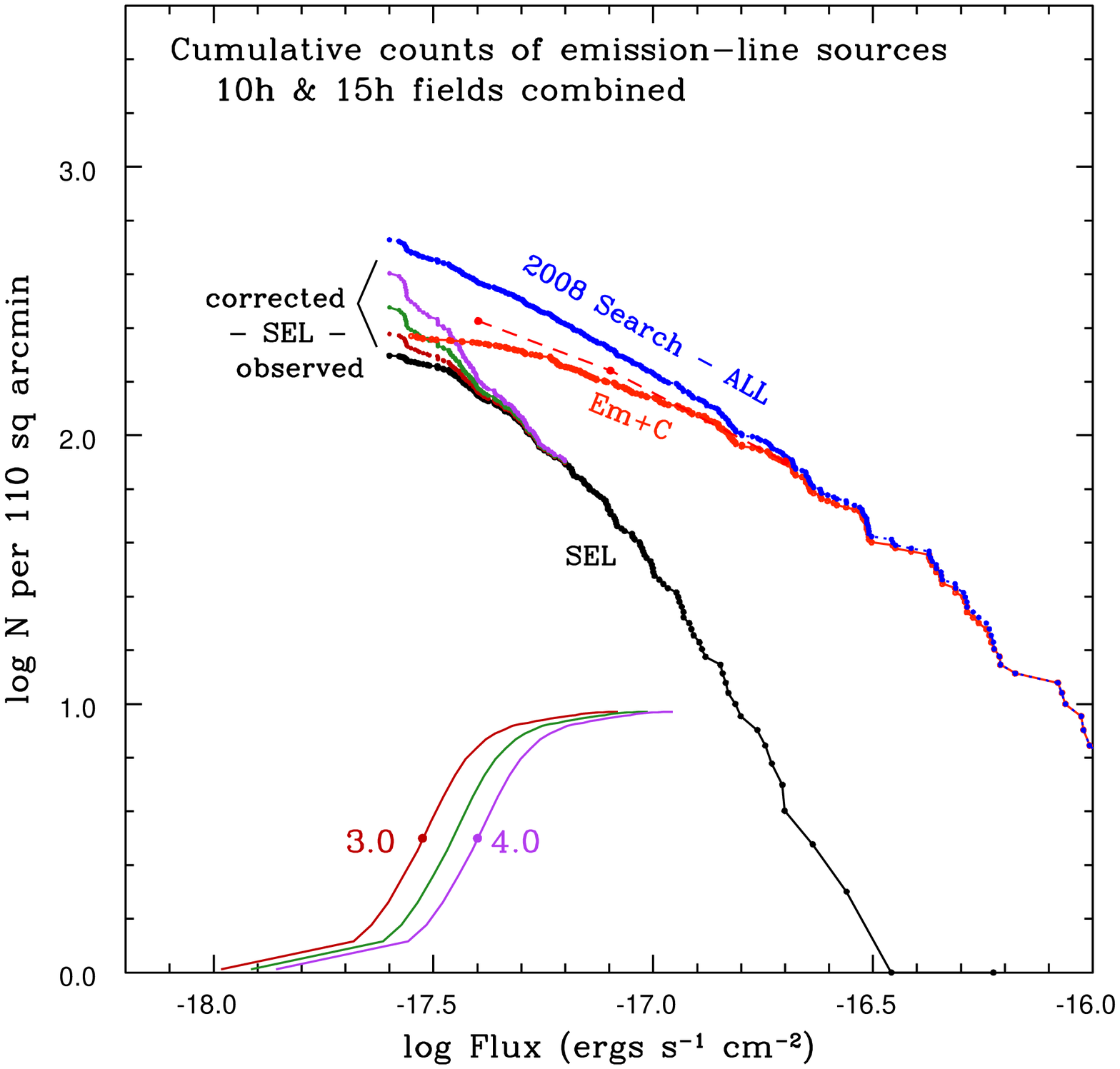,angle=0.,width=3.4in}}
\noindent{\scriptsize
\addtolength{\baselineskip}{-3pt}
 
\hspace*{0.3cm}Fig.~4.\ The cumulative source counts formed for both the 10h (COSMOS) and 15h (LCIRS)
fields, in log$_{10}$ number per 110 sq arcmin, the combined area surveyed in the two fields. The red line shows the 
cumulative counts of `Em+C' (emission-line sources with blue continua); the black line shows the counts of `SEL' 
(single-emission-line-only) sources.  The faint end of the latter is shown in raw counts as well as after correction for three 
different values of the completeness correction (see text), at 50\% incompleteness levels of  $F = 3.0, 3.5, $ and $ 
4.0\, \10-18$\flux. (The form of the incompleteness function is shown at the bottom of the figure, and the correction
applied to the observed counts -- the black line -- is shown in red, green, and purple for the these three bounding values
of the 50\% completeness level.)  The red-line Em+C distribution has not been corrected for this incompleteness, but 
for a larger effect that results from the increasing difficulty of detecting a continuum for progressively fainter emission 
lines.  The dashed red line is an estimate of the correction that is needed because of this effect.  As explained
in the text, the `ALL' line, a sum of the uncorrected Em+C counts and the SEL counts corrected for the `3.5' 
incompleteness, has instead been chosen as the basis for the subsequent analysis of the LAE luminosity function.   

\vspace*{0.2cm}
\addtolength{\baselineskip}{3pt}
}

As can be appreciated from the sharp peak in the emission-line-only sources in Figures 1 \& 2, incompleteness 
in our SEL detections sets in sharply at $F \approx 3\,\10-18$\flux.  (This is seen in the cumulative counts of 
Figure 4 as the sudden leveling-off of the black line.)  In Martin08 we investigated incompleteness by
the technique of randomly inserting point sources of different flux levels and using SExtractor to try to recover
these sources.  The result of that experiment was just such a sharp cutoff, whose form is reproduced in Figure 4
for three different flux values of 50\% incompleteness.  When this form of the incompleteness correction
is applied for 50\% flux values of 3.0, 3.5, 4.0 $\10-18$\flux, we obtain the red, green, and purple extensions
to the black line of Figure 4.  The result is very reasonable, extending the SEL source counts at roughly 
the same slope down to the limit of our detections, and constraining the 50\% point to a small range.  For the 
2008 MNS search data, then, we adopt the Martin08 form and choose $F = 3.5\,\10-18$ as the flux where 
incompleteness falls to 50\%.  We emphasize that we use the incompleteness correction only to demonstrate that
there is no evidence for a `turnover' in the SEL counts to the limit of our observations.  Because the steep slope of 
the black line is well established before the small interval over which the incompleteness correction is applied, 
none of the results in this paper rely on applying this correction.

The incompleteness correction has not been applied to the Em+C source counts. (In fact, so few galaxies 
are being added to the Em+C counts at the limit of the survey that making the correction used for the 
SEL counts would add only about 7 sources.)  A more important incompleteness, one that sets 
in more gradually, comes from selecting the Em+C sources with a combination of continuum and line flux. 
The issue is that increasing numbers of the faintest foreground sources will not be identified as such if their 
stellar continua are too faint to be detected in our survey.  These sources will not be lost, but will be added
to the SEL counts if their emission-line fluxes are above the detection threshold.  In order to estimate the 
magnitude of this problem, we measured equivalent width (EQW) of 212 emission lines found in the Em+C 
sample and divided the sample into three flux  ranges,  $4 < F < 8$ (67), 8 < F < 20 (76), and 20 < F < 1000 (69), 
where F is the flux in units of $10^{-18}$\flux\ (sample size in parentheses).  We assume the brightest sample 
represents an intrinsic  EQW distribution that is independent  of emission-line flux.   As expected, the two fainter 
distributions show a deficiency of higher equivalent widths: the median equivalent width drops from 66$\ang$ 
for the brightest sample to 43$\ang$ for the intermediate flux sample to 21$\ang$ for the faint sample.  From the 
faintest sample it is clear that incompleteness sets in at $EQW \gs 50\ang$; there are essentially no 
$EQW > 100\ang$ cases.  Taking the conservative case of $EQW = 50\ang$ as the detection limit for the medium 
and faint samples, we calculate that 47 sources are missing from the Em+C counts;  presumably, these have 
been added to the SEL sample.  This simple modeling suggests that approximately 22\% of the SEL sample are 
lower-flux foreground objects whose continuum is too faint to be detected.  While significant, this is not a  large 
fraction of SELs, so there must be higher equivalent width foreground sources (EQW $\gs 500$ --- missed even for 
our brightest sources) and/or genuine no-continuum sources, for example, \Lya\ emitters, that dominate dominate the 
SEL sample.

In Figure 4 we show the Em+C line as corrected for EQW incompleteness as the dashed red line. Because it 
unlikely that the equivalent-width distribution is independent of luminosity, as we have been obliged to assume, this 
correction is not accurate enough to separate the LAE population from foreground galaxies.
Therefore, we will not use the red line Em+C distribution, in raw or corrected form in the analysis
that follows.  Instead, we combine the uncorrected Em+C distribution with the SEL sources (corrected for 
incompleteness with 50\% incompleteness at $F = 3.5\,\10-18$\flux) --- this is the `ALL' emission-line sources seen 
in Figure 4 (the blue line).  With this we will circumvent the uncertainty associated with detection of a faint 
continuum, by subtracting an independently derived foreground population from the combination of LAEs 
and foreground galaxies that together comprise the full MNS sample of emission-line sources.

\section{Analysis: Is the steep rise in faint SEL sources due to LAEs at z = 5.7?}

Although it is certain that some of the rising SEL counts are LAEs at $z = 5.7$, the size and significance of this 
population depends on how many of these sources are foreground galaxies.  A decisive answer to this question 
will come from higher resolution spectra of the fainter LAE candidates to distinguish, in particular, \Lya\ emission
from \OII\ --- the common foreground contaminant that is most likely to appear as a single line at the spectral resolution 
of the 2008 search data.  With a spectral resolution of a few angstroms, \Lya\ will often exhibit an asymmetric profile, 
sharply attenuated to the blue but with a red ``wing,'' while the \OII\ doublet lines will be resolved 
($\Delta \lambda =  5.7$\ang\ for $z = 1.19$), thus providing a definitive test of the two most likely possibilities.  The 
challenging task of obtaining a statistically significant sample of higher resolution spectroscopy for our fainter 
candidates is underway at the Keck and Magellan telescopes.

\subsection{Using the COSMOS data to extrapolate foreground contamination}

In lieu of decisive spectroscopic confirmation of faint LAEs, we explore what can be learned 
by a statistical subtraction of the foreground population.   Our own observations of foreground Em+C galaxies, 
like our sample of SEL sources, is the faintest such sample available.  However, as discussed above, our data 
themselves show that the faintest foreground sources are not identified as such when the stellar continuum 
is too weak to be detected.  
 
To work around the problem --- that the lack of a detected continuum for an SEL does not necessarily imply that this object
is an LAE --- we take advantage of a wider-area search for foreground emission-line galaxies by the COSMOS 
Survey, observations with Suprime-Cam that include numerous broad bands as well as intermediate- and narrow-band 
filters (Taniguchi \etal\ 2007).  One in particular, the NB816 filter, covers a bandpass close to that used in our study: central 
wavelength $\lambda_c = 8150$\ang\ for NB816 compared to our $\lambda_c = 8185$\ang, and FWHM bandwidth 
$\Delta\lambda = 121$\ang\ compared to our $\Delta\lambda = 134$\ang.  Combining narrow-band detections of 
emission-line sources with detections in several broad bands allows good discrimination into different redshift intervals, 
as described in Takahashi \etal\ (2007) and Shioya \etal\ (2008) (see also Ly \etal\ 2007).   This much wider-area search 
for emission-line sources provides a sample of foreground galaxies that is populous enough to well constrain the luminosity 
function.  The COSMOS observations reach a limiting flux $F = 10^{-16.85}$\flux, although --- as discussed in 
\S2 --- approaching this flux limit the uncertainty in these detections rises considerably.   Nevertheless, the 
samples are large and uniform; they provide a solid database for determining foreground luminosity 
functions that can be extrapolated to the flux limit of our LAE search.


\vspace*{0.1cm}
\hbox{~}
\centerline{\psfig{file=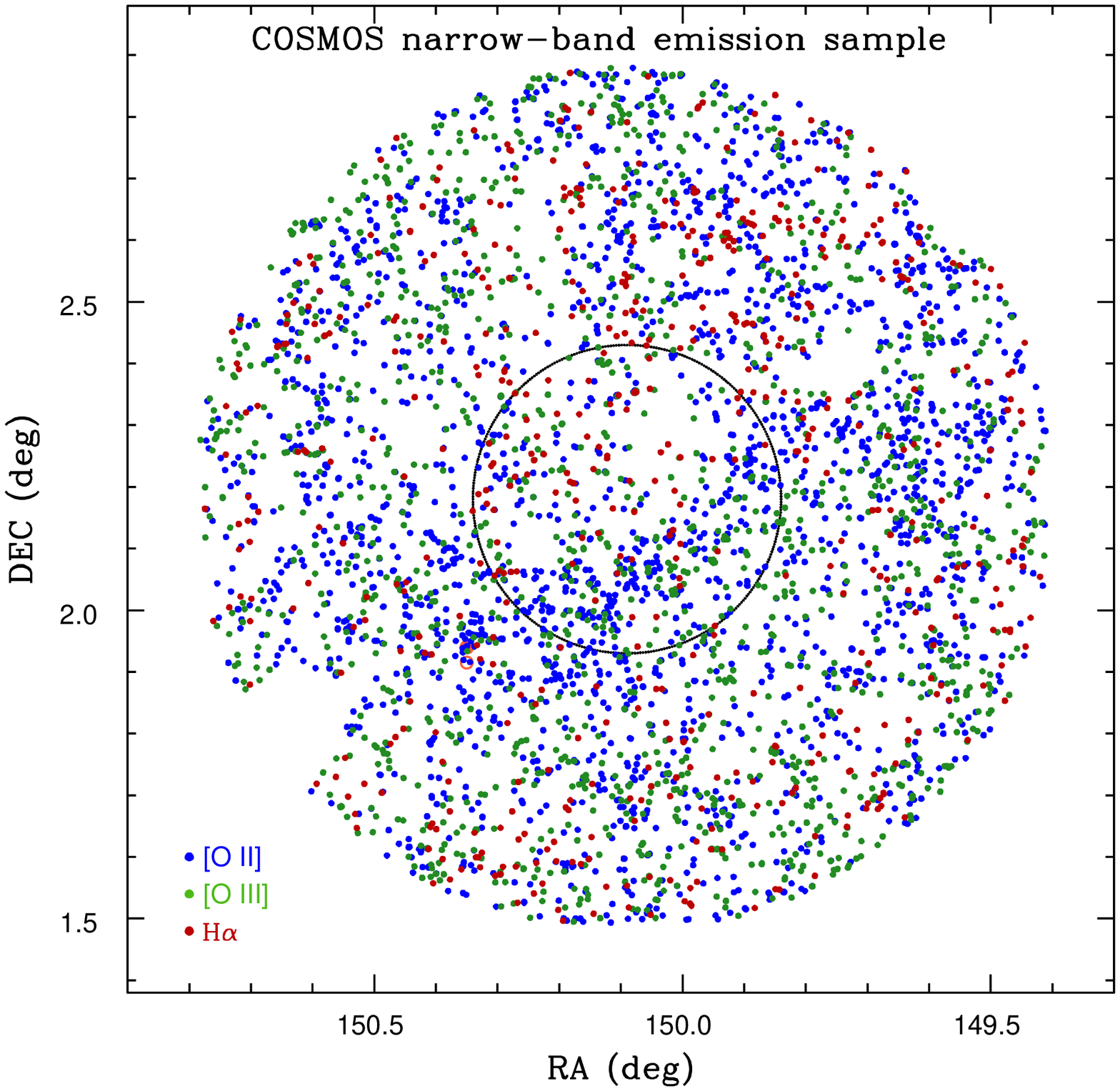,angle=0.,width=3.3in}}
\noindent{\scriptsize
\addtolength{\baselineskip}{-3pt}
 
\hspace*{0.3cm}Fig.~5.\ The COSMOS survey detections of emission-line galaxies with a blue continuum in the NB816 
filter, which includes emission-line sources brighter than $F = 10^{-16.85}$\flux.  Components of this sample are \OII\ 
emission at $z \approx 1.19$, \OIII\ at $z \approx 0.63$, and \Ha\ at $z \approx 0.25$.  The smaller circle is the \emph{IMACS} 
field of our 2008 search; the larger circle, with a 0.70\deg\ radius, is the area we used for determining the foreground 
contamination per area. (This is most but not all of the rectangular area of the NB816 catalog.) Large-scale structure 
in the distribution of foreground galaxies can be clearly seen, particularly for the \OII\ emitters.

\vspace*{0.2cm}
\addtolength{\baselineskip}{3pt}
}

Figure 5 shows the distribution of emission-line sources also detected in continuum bands, parsed into \OII\ at 
$z \approx 1.19$, \OIII\ at $z \approx  0.63$, and \Ha\ at $z \approx 0.25$.\footnote{In fact, no broad-band
photometric criteria for selecting \OIII\ explicitly has been published in the COSMOS study.  We have therefore 
taken the \OIII\ sample to be what remains of the NB816-excess ($\Delta$mag $> 0.20$) sample after removing 
the \OII\ and \Ha\ sources, which means there is also a smaller contribution of H$\beta$\ emitters ($z \approx 0.68$) 
in the \OIII\ sample.   The \Ha sample is likely to contain a similar contribution of [S II] sources.} For a sample 
this bright, almost all emission-line sources belong to one of these foreground populations.  These data are 
extracted from the COSMOS Intermediate and Broadband Photometry 
Catalog;\footnote{http://irsa.ipac.caltech.edu/Missions/cosmos.html} emission-line galaxies are selected following the
narrowband excess method described in Takahashi \etal\ (2007).  This sample provides the crucial advantage for our 
purposes that fluxes have \emph{not} been corrected for internal extinction --- correcting for extinction is commonly 
done because most studies have focused on the star formation rate over cosmic time

The sample we extracted is centered on our 2008 search field, shown by the inner circle in Figure 5, and extending 
to a radius of 0.7\deg --- most, but not all, of the area of the COSMOS NB816-excess catalog.  The percentage 
compositions of \OII, \OIII, \& \Ha\ for the larger area are 53\%, 34\%, and 13\% respectively, and 55\%, 32\% and 
13\% for the smaller area of the 2008 search.  Figure 5 shows that \OII\ at $z \approx 1.19$ is the major foreground 
and that, for these \OII sources in particular, there is apparent large-scale structure --- a prominent filament 
structure crosses the \emph{IMACS} field.  (The COSMOS photometric redshifts place this filament at 
$1.18 < z < 1.20$, within the bandpass of both the NB816 and our MNS narrow-band filter.)  This feature offers 
the possibility of a simple test comparing the spatial distributions of LAE candidates and foreground emitters to 
estimate the degree of foreground contamination.

\subsection{Evidence from cross-correlation that many of the faint sources are LAEs}

The relatively strong filament of foreground \OII\ emitters, the most numerous of the foreground sources, suggests 
that we might cross-correlate positions of \OII\ sources in the COSMOS catalog to estimate what fraction of the 
candidate LAEs are instead members of this foreground.   Although our sample of LAE candidates extends somewhat 
fainter, there is substantial overlap in the fluxes of the two samples.   

We have made such a test by constructing angular correlation functions, various forms of which are discussed by 
Landy \& Szalay (1993).  In our particular application we cross-correlate the spatial positions of the SEL sources --- candidate 
LAEs --- with the largest foreground population, the sample of \OII\ sources at $z \approx 1.19$ from the COSMOS survey.  
Specifically, we measure the angular separation in arcminutes for all SEL--\OII\ pairs and compare to expectations for 
randomized placements of the SEL sample over the search field. In the notation of Landy \& Szalay, the angular 
cross-correlation function we use is $w(\theta) = (D_{SEL}D_{[O II]}/R_{SEL}D_{[O II]}) - 1$.


\vspace*{0.1cm}
\hbox{~}
\centerline{\psfig{file=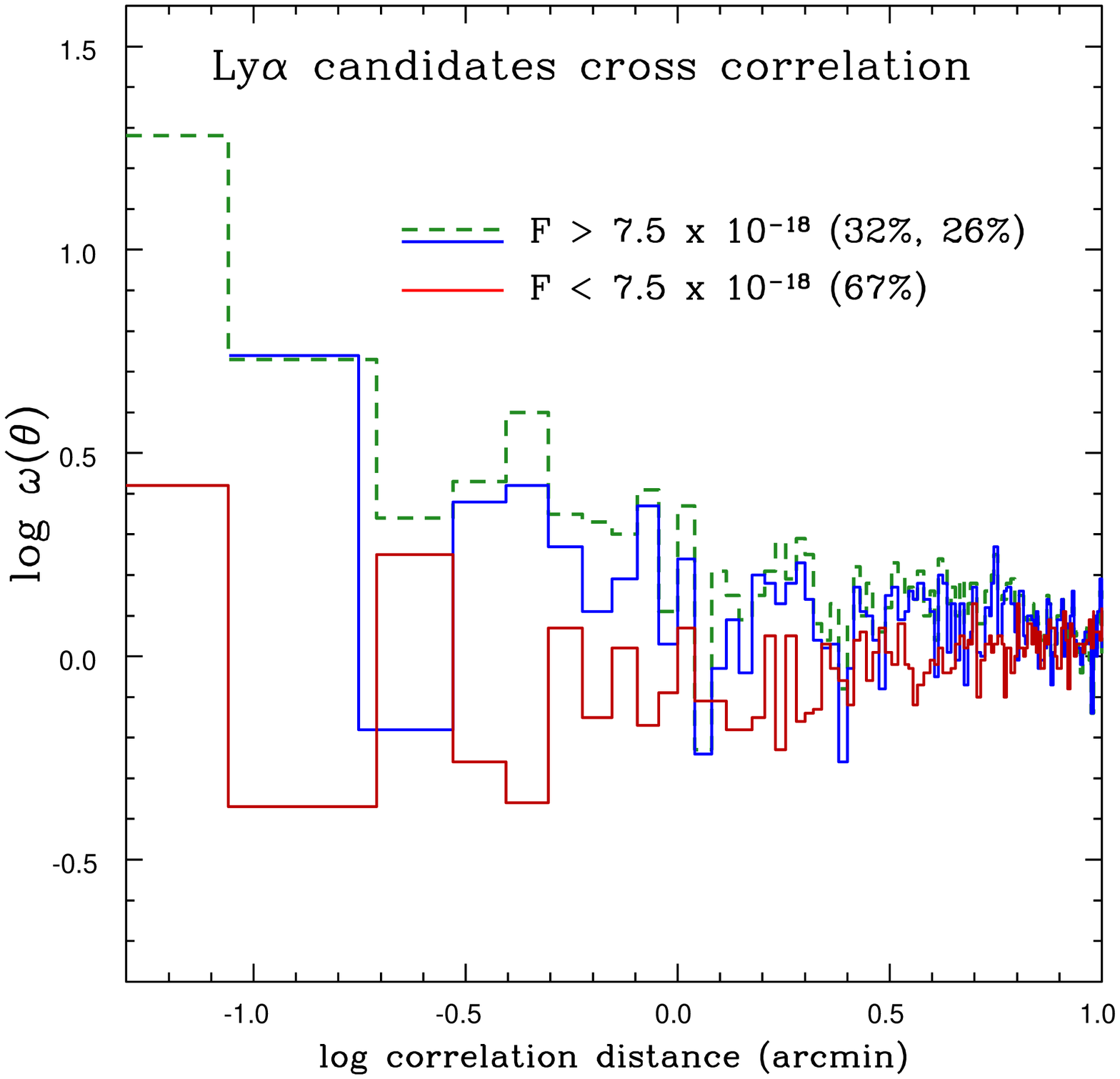,angle=0.,width=3.2in}}
\noindent{\scriptsize
\addtolength{\baselineskip}{-3pt}
 
\hspace*{0.3cm}Fig.~6.\ The angular cross-correlation function $\omega(\theta)$ between SEL sources found in our 
2008 search of the 10h field and the COSMOS narrow-band detections of galaxies with \OII\ emission at 
$z \approx 1.19$ overthe full \emph{IMACS} field.  The dashed green line shows a very strong correlation between 
33 bright SEL sources ($F > 7.5\,\10-18$\flux) and foreground \OII\ emitters, however, the strong signal in the first bin
is due to the fact that 6 of these bright SEL sources are certain or probable HII regions in galaxies in the \OII\ 
sample.  However, even without these, the correlation function for the bright sample --- the blue line -- is still 
strong, a $4\sigma$ or greater difference from `random.'   In contrast,  the red line shows that the 70 faint SEL 
sources ($F <  7.5\,\10-18$\flux) are uncorrelated with the foreground \OII\ emitters, suggesting the latter 
make up half-or-less of this population.  Monte Carlo models show this result to be very significant as well, although 
the unavailability of a foreground population as faint as our sample prevents a good constraint on how big the 
foreground contribution can be.  This decline in amplitude of the cross-correlation function with decreasing flux suggests 
that the steep rise in SEL sources is the result of a large population of LAEs at $z=5.7$.

\vspace*{0.2cm}
\addtolength{\baselineskip}{3pt}
}

The results of the cross correlation test are shown in Figure 6.  We know from Martin08 and from the SDF
search for LAEs (Shimasaku \etal\ 2006) that foreground dominates over LAEs by $\sim5-10:1$ for a 
sample with $F \gs 10^{-17}$\flux.  Therefore, we divided the SEL sample of LAE candidates at $F = 7.5 \10-18$ 
\flux\ --- 33 sources are brighter, and 71 are fainter than this value.  The dashed green line in Figure 6 shows the 
result for the bright sample --- a very strong correlation signal over the 6 bins of pair separations $R \le 0.6$\arcmin.  
Over this interval 46 pairs are observed, compared to only 13 predicted for a random placement of the \OII\ 
sources --- statistically significant at more than $10\sigma$.  However, it is clear that this result is strongly 
driven by the smallest pair separations $\delta \le 0.1$\arcmin, where there are 7 pairs from 6 galaxies.  
Suspiciously, the largest of the separations was 1.8\arcsec, much smaller than the $\delta \le 6$\arcsec\ 
covered by this bin.  After examining these with the HST-ACS images of the COSMOS field, we recognized 
that --- as found in Martin08 --- 3 cases were certainly members (and the other 3 probable members) of the 
\OII\ catalog with which we were correlating, that is, these SELs appear to be HII regions in the outer parts 
of foreground galaxies.  On this basis we made a conservative choice to remove all 6 sources from the bright 
sample, leaving 27.  Significantly, only one such close pair ($\delta = 1.0\arcsec$) was found for the 71 faint 
SEL sources (even fainter HII region of the same foreground galaxies should be detectable), which was 
similarly removed. 
  
Figure 6 shows that the angular cross-corrleation signal for the remaining bright sample of 27 remains significant, indicating 
that many of these SEL sources are foreground members at $z \approx 1.19$.  With the central bin eliminated, there is still 
a clear signal --- the blue line -- for the 5 remaining bins  that tally 24 pair separations up to 0.6\arcmin. Over  this interval, 
only 11 pairs are predicted for a random spatial distribution of the 27 sources --- a $\sim7\sigma$ difference.  
In contrast, for the remaining 70 SELs with $F < 7.5\,\10-18$\flux, there is no clear correlation signal over the same 6 
bins: 28 are expected and 26 are found --- this null result is also very significant, as we now demonstrate.

Monte Carlo simulations can be used to enhance this analysis and, in particular, to test whether the strength of the correlation 
or lack thereof is consistent with expectations.  In our simulations we constructed simulated SEL samples of 27 and 70 by 
drawing randomly from the \OII\ population and adding a number of randomly placed objects to simulate other objects.  
For example, we expect the majority of the SEL sample with $F > 7.5~\10-18$\flux\ to be foreground galaxies, and that 
roughly half of these should be \OII\ emitters.  Specifically, we tested for the 27 bright sources by making 1000 realizations in 
which we extracted 14 objects from the \OII\ sample (removing them temporarily from that sample) and also placed 13 other 
objects at random positions to represent --- in the proportions of the COSMOS narrow-band sources --- 9 \OIII and 2 \Ha\ --- this 
simulation produces a mean number of 24 separations with $R \le 0.6$\arcmin,  the same as the 24 pairs of the observed sample 
and leaves room for 2 LAEs at $z = 5.7$.  However, because of the small sample size, the simulations cannot well constrain the 
foreground or LAE fraction.  For example,  a mix of 8 \OII\ selections (plus 6 other foregrounds and 13 genuine LAEs) will also yield 
24 pair separations $R \le 0.6$\arcmin\ at the $1\sigma$ level.  Nonetheless, the case of \emph{zero} foreground produces only 4 out 
of 1000 models with 24 or more pairs, a $>3\sigma$ rejection that is at least qualitatively in agreement with the simple Poisson 
calculation done above.  In summary, the bright 33 SEL sample is dominated by foreground objects, 6 that are identified as actual 
COSMOS sources, and --- from the Monte Carlo models --- between 58\% (the $1\sigma$ lower limit) and 100\% of the 
remaining 27 sources also foreground.

The Monte Carlo simulations are also effective in testing the significance of the null signal for the 70 candidate sample 
with $F < 7.5~\10-18$\flux.  Adopting the same fractions as the `best fit' case above --- 36 \OII\ and 34 random (\OIII, \Ha, 
and LAE) --- leads to a mean expectation value of 60 pairs for $R \le 0.6\arcmin$, far exceeding the measured value of 26 for 
the observed 70 SEL sample -- this `nearly all foreground case' appears to be ruled out at the $\sim4\sigma$ level.  The 
`half-foreground' case, 19 \OII\ and 16 \OII\ and \Ha, with a 50\% population of LAEs, has a mean expected value to 
46 --- still unacceptably large, although `26 pairs' is allowed at the $2\sigma$ level.  To reach a $1\sigma$ level the 
foreground must be reduced to 7 \OII\ draws and 6 random placements.  Taken at face value, then, the lack of a 
prominent signal in Figure 6 for the 70 faint source sample of candidate LAEs suggests that the foreground population 
should be less than 25\%.  This further suggests a very sharp change in foreground contamination which, anticipating 
the result of \S5.4, would require an improbably steep rise in LAEs sources.

There is, however, a further issue to consider.  The 27 source bright sample covers the same flux range 
as the COSMOS \OII\ catalog, making a simple draw from the \OII\ distribution a reliable way to test the expected strength of
the correlation function.  In contrast, the faint sample of 70 sources are all fainter than anything in the COSMOS catalog.  
In general, correlation functions are observed to depend on source luminosity (or mass) in the sense that higher luminosity 
sources are more strongly clustered.  We tested the likely strength of this effect by dividing the \OII\ sample in half, split at 
log $F$ (ergs s$^{-1}$ cm$^{-2}$) $= -16.57$. Returning to the `half foreground' case, when we draw only from the brighter 
half of the \OII\ sample, the Monte Carlo simulation predicts a mean expected number of 49 \OII\ sources, while drawing 
from the faint sample the mean number drops to 40 --- a significant weakening of the correlation strength.  If this trend 
continues to the yet-fainter 70 source sample, the `half-foreground' sample would be compatible with the weak 
cross-correlation signal between the faint sample of LAE candidates and the COSMOS \OII\ foreground.  While our 
samples of candidate LAEs are too small to accurately quantify this effect, its sign is clear.  Unfortunately, this 
uncertainty makes it difficult to constrain the foreground contamination within the 0-50\% level, but the `nearly all 
foreground' case found for the bright sample remains highly unlikely.  Recall also that only a single pair was removed 
from the sample of 70 faint sources as a likely HII region of the foreground \OII-emitting 
population.

In summary, when the sample of 104 SELs in the 10h search field is divided into brighter/fainter than $F = 7.5~\10-18$\flux, the 
difference in strength of the angular cross-correlation with the [O II]-emitting foreground is highly significant.  The strong signal 
observed for the brighter 33 SEL sample, even after removing 6 actual foreground objects, indicates that the majority of this 
sample is foreground and only a minority can be LAEs --- consistent with previous studies.  In contrast, the absence of a 
strong signal for the faint 70 SEL sample rules out the possibility that the most are foreground galaxies, indicating instead 
that approximately half are LAEs.  This is our first good evidence that the rapid rise in the faint SEL counts is due to a 
substantial increase in the number of LAE emitters at $z=5.7$, something we can investigate further by looking at the 
luminosity functions of the foreground sources, information that is independent of their positions on the sky.

\subsection{Fitting Schechter luminosity functions to the COSMOS foreground counts}

With good measurements of the observed foreground populations of \OII, \OIII, and \Ha\ emitters
with fluxes $F > 10^{-17}$\flux\ from the COSMOS Survey, we proceed to use these counts to
predict the foreground population down to the faintest levels of our survey, $F \sim 2.5\,\10-18$\flux.
This was accomplished by binning the observed counts and fitting Schechter functions.  In general, 
this was a straighforward procedure, except for the case of \OII, which makes up more than 
half of the foreground.  We found that the best $\chi^2$ fit of the \OII\ flux distribution skewed 
the fit of the faint end slope --- critical for our purposes here --- to a steeper slope, $\alpha = -1.72$, 
than the data.  To make a better fit of the faint end slope, we used a one-sided Kolmogorov-Smirnov 
(KS) test for goodness-of-fit to find the best fitting slope.   Consistent with what we had observed, 
this test ruled out this best $\chi^2$ fit at the 97\% confidence level, as described below.

The KS test works with cumulative distributions.  For this reason it is well suited to our application, since 
it is the cumulative source counts shown in Figure 4 that we want to model.  The one-sided KS test
can be used to compare a model distribution, in this case an integrated Schechter luminosity function,
to observed data, here, cumulative \OII\ counts; we used it to find the most probable 
faint-end slope and its allowable range.  Both distributions were normalized to the interval 0.0 to 1.0 and 
the maximum fractional deviation and total count (2323 \OII\ sources) was entered into an online KS 
calculator\footnote{http://www.ciphersbyritter.com/JAVASCRP/NORMCHIK.HTM\#KolSmirof} that returned
the probability that the data were drawn from a distribution following the Schechter function.  Since the 
two distributions are normalized, the only free parameter is the characteristic luminosity $L^*$. 
For each value of slope $\alpha$ we found the value of $L^*$ that produces the minimum fractional 
deviation in the KS test.


\vspace*{0.1cm}
\hbox{~}
\centerline{\psfig{file=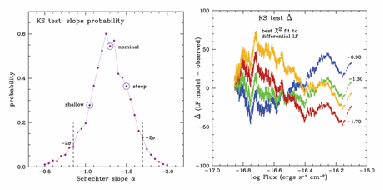,angle=0.,width=3.9in}}
\noindent{\scriptsize
\addtolength{\baselineskip}{-3pt}
 
\hspace*{0.3cm}Fig.~7.\ KS test results of fitting Schechter functions to the \OII\ counts. (left) The probability of fit for 
different values of the slope $\alpha$ and luminosity $L^*$.  The two vertical dashed lines mark the slopes of 
$\alpha = -0.85$ and $\alpha = -1.70$, representing fits that are rejected at the $\sim95\%$ level ($2\sigma$). 
(right) Four examples of the fractional deviations $\Delta$ for best-fit Schechter functions and the \OII\ counts.  
The larger the amplitude of the S-wave shape of  the $\Delta$ distribution, the poorer the fit.   The best fit, $\alpha = -1.30$, 
is nearly flat --- well within the $1\sigma$ interval of  ``goodness of fit'' of -1.05 to -1.50   The best $\chi^2$ fit, shown in gold, 
shows a stronger S-wave that indicates rejection at the $P\, > $ 97\% level.

\vspace*{0.2cm}
\addtolength{\baselineskip}{3pt}
}

In Figure 7 we plot the  probability that the \OII\ data are drawn from a Schechter model of slope $\alpha$, 
for the value of $L^*$ that maximizes this probability.  The best fitting slope is $\alpha = -1.30$ and the
equivalent of a $1\sigma$ interval is -1.05 to -1.50.  What is in effect a $2\sigma$ rejection of the
model occurs at $\alpha = -0.85$ and $\alpha = -1.70$.  Figure 7 also shows the integrated deviations 
(in counts) for the four Schechter function fits of the data, including the best $\chi^2$ fit for the \OII\
LF with faint-end slope of $\alpha = -1.72$.  Though the residuals of this latter case are not large 
compared to Poisson errors, the S-wave shows a systematic error in the fit that accounts for the 
anomalously high value of $\alpha$.

Choosing the KS test to find the best fitting slope makes sense because our application is insensitive to 
how well the \OII\ distribution fits a Schechter function around $L^*$.  If a Schechter function was known to be an 
excellent parameterization (as it is for broad-band galaxy fluxes) for galaxy emission-line luminosities \emph{uncorrected 
for extinction}, it could be argued that the faint end extrapolation would be best made by giving as much weight to the 
curvature around $L^*$ as to the power law faint end.  However, since the effects of extinction could subtly alter
the form from a Schechter function, we believe that in this case a fit that gives more weight to the faint-end slope is preferred. 

The adopted Schechter-function fits are compared to the binned differential counts in Figure 8.  The values of the
parameters ($\alpha, L^*, \phi^*$) we will use for the rest of this analysis are also given in the figure.  In particular, 
the slopes we have derived are consistent with the values derived by other studies that made Schechter-function fits 
of foreground populations at these redshifts.  For example, for \OII, \OIII, and \Ha\ slopes, respectively, Hippelein
 \etal\ (2003) find values of -1.45, -1.50, -1.35, and Ly \etal\ (2007) find values of -1.15, -1.22, -1.70.  For the COSMOS 
survey itself, Takahashi \etal\ (2007) report an \OII\ slope of -1.41 and Shioya \etal\ (2008) report an \Ha\ slope of -1.35.


\vspace*{0.1cm}
\hbox{~}
\centerline{\psfig{file=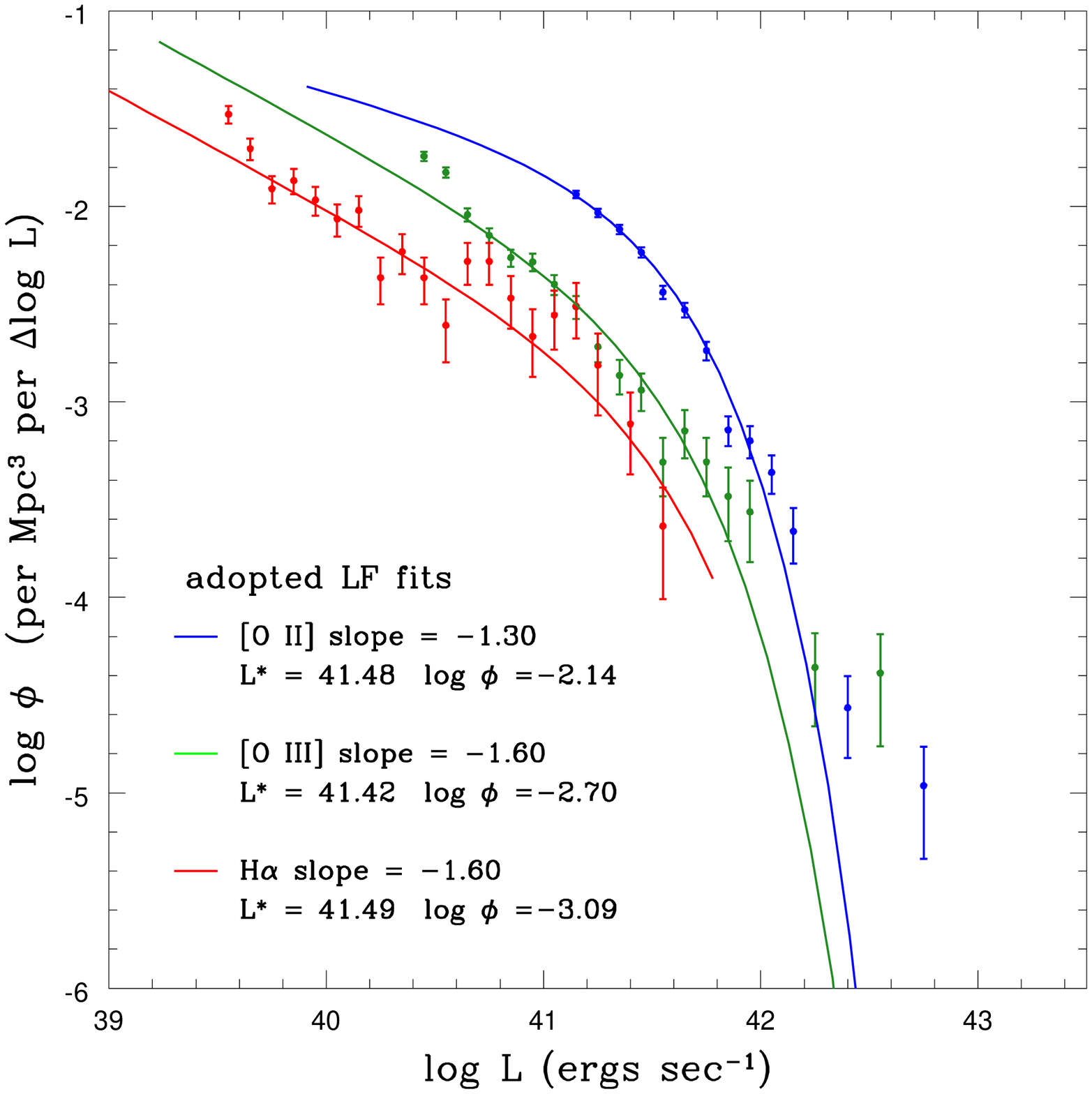,angle=0.,width=3.4in}}
\noindent{\scriptsize
\addtolength{\baselineskip}{-3pt}
 
\hspace*{0.3cm}Fig.~8.\ Differential LFs for sources in a 0.7\deg\ radius circle surrounding the search field position.
There are 2323 sources in the \OII sample, 1491 \OIII sources, and 600 \Ha sources within the 0.7\deg\ circle 
that have been used for fitting luminosity functions (see text).  Slope values of -1.30, -1.60, and -1.60, respectively, 
have been used to extrapolate the expected number of foreground galaxies down to $F = 2.5\,\10-18$\flux. The
corresponding values of $L^*$ in ergs s$^{-1}$ and log\,$\Phi$ in Mpc$^{-3} (\Delta$logL)$^{-1}$ are recorded.

\vspace*{0.2cm}
\addtolength{\baselineskip}{3pt}
}

\subsection{Subtracting extrapolated foreground populations to measure the LAE luminosity function }

We have used these LFs and their extrapolations $2\,\10-18 < F < 10^{-17}$\flux\ to predict forgeround contamination 
down to the limit of our survey.  In Figure 9 we compare the counts of our LAE search (shown in Figure 4) to the 
the observed COSMOS foreground counts (blue, green, and red lines for \OII, \OIII, and H$\alpha$) and the Schechter
function fits to these counts (solid magenta lines, then dotted for the extrapolation).  Though we have used all the 
data within the 0.70\deg\ circle --- 2323 \OII, 1491 \OIII, and 600 \Ha sources --- to fit the LFs, we normalize to the counts
for these foregrounds in the \emph{IMACS} search field, 358, 201, and 81, respectively, for a total of 640.  The summed result 
is the dashed green line marked `COSMOS' in Figure 9. This summed COSMOS foreground is a very good fit to the 
`Em+C' counts of our LAE search --- the red line, down to $F = 10^{-17}$\flux.  Fainter than this the \emph{extrapolated} 
COSMOS foreground falls between the certain foreground (red) and total observed (blue), suggesting that while some 
of the SEL sources are foreground, many are plausibly identified as LAEs.  Using these nominal slopes, then, typical of 
those measured for comparable galaxy populations at $z \ls 1$, leads to the conclusion that the  steep slope of SEL 
sources is due in part to an increasing fraction of LAEs.


\vspace*{0.1cm}
\hbox{~}
\centerline{\psfig{file=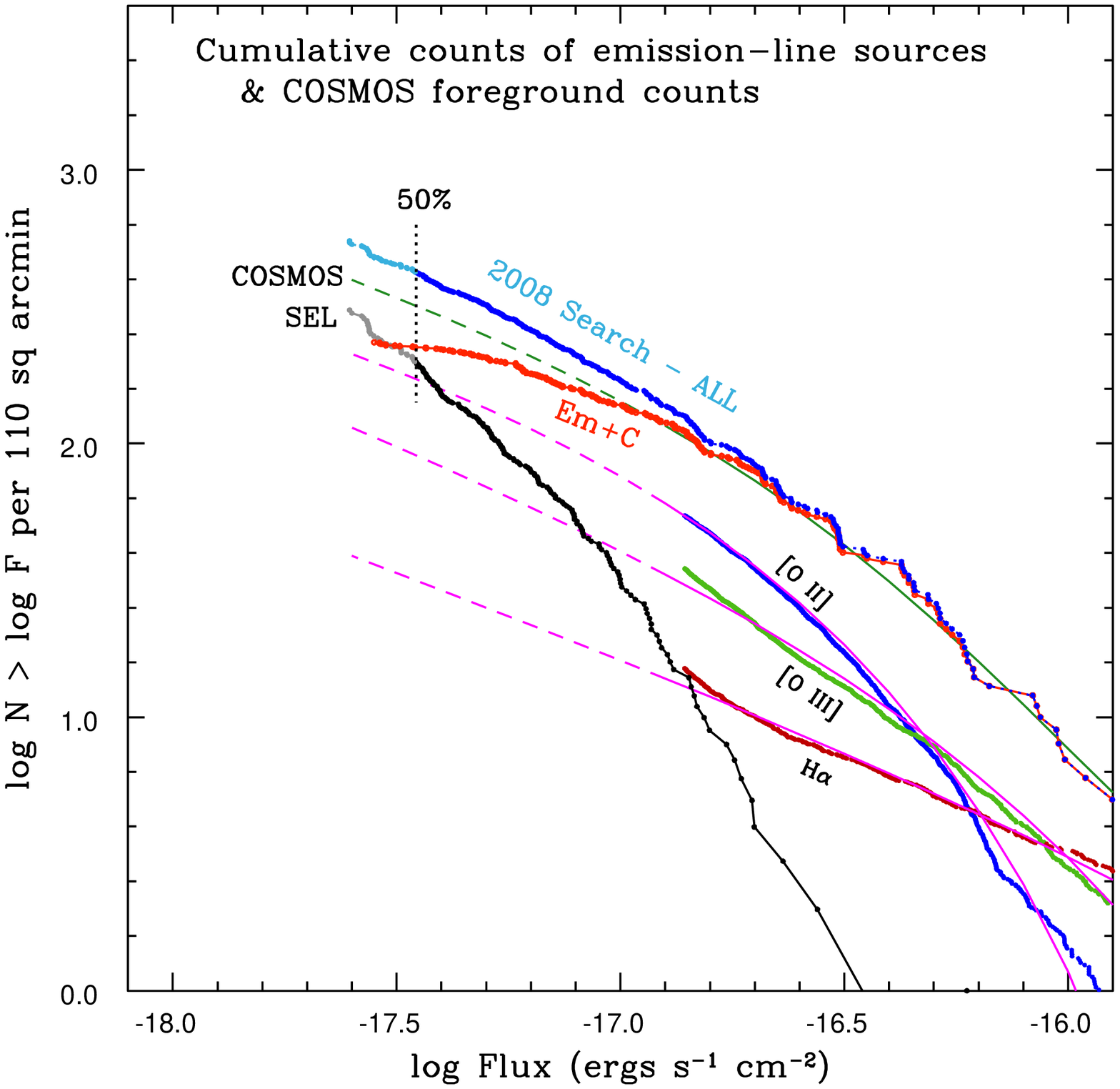,angle=0.,width=3.2in}}
\noindent{\scriptsize
\addtolength{\baselineskip}{-3pt}
 
\hspace*{0.3cm}Fig.~9.\ The same as Figure 4, but including cumulative counts of \Ha, [O III], and [O II] foregrounds from 
the COSMOS survey catalog of Takahashi \etal (2007), with fits to Schechter functions made by us for extrapolation fainter than 
$F = 10^{-17}$\flux.  As before, the `ALL' line is the sum of SEL and Em+C sources.  (The adopted incompleteness 
correction described in \S4 has been applied.  The vertical bar marks the 50\% incompleteness at $F = 3.5\,\10-18$\flux: 
at this point the cumulative `ALL' counts are greater by 7.0\% than for an uncorrected distribution.)   The `COSMOS' line is the 
sum of these three expected foreground interlopers, as explained in the text.  Since the `ALL' line of our 2008 survey is a sum of
foreground sources and LAEs at $ z = 5.7$, if the extrapolated foreground of the COSMOS data accurately represents all foreground 
objects, then its difference from the `ALL' line of our search should be due entirely to LAEs.

\vspace*{0.2cm}
\addtolength{\baselineskip}{3pt}
}

We quantify this in Figure 10, which shows the cumulative luminosity function (LF) of LAEs obtained by subtracting the extrapolated 
COSMOS foreground. The upper blue and green lines (identical to those in Figure 9) correspond to the total observed counts (corrected 
for incompleteness) and the predicted foreground counts from fits to the COSMOS NB816 data.  The difference ---`observed' minus 
`foreground' --- is shown as the magenta line and compared to the {\it extrapolation} of the fits to the LF for LAEs in the 
Subara Deep Field (SDF, Shimasaku \etal\ 2006).  


\vspace*{0.1cm}
\hbox{~}

\centerline{\psfig{file=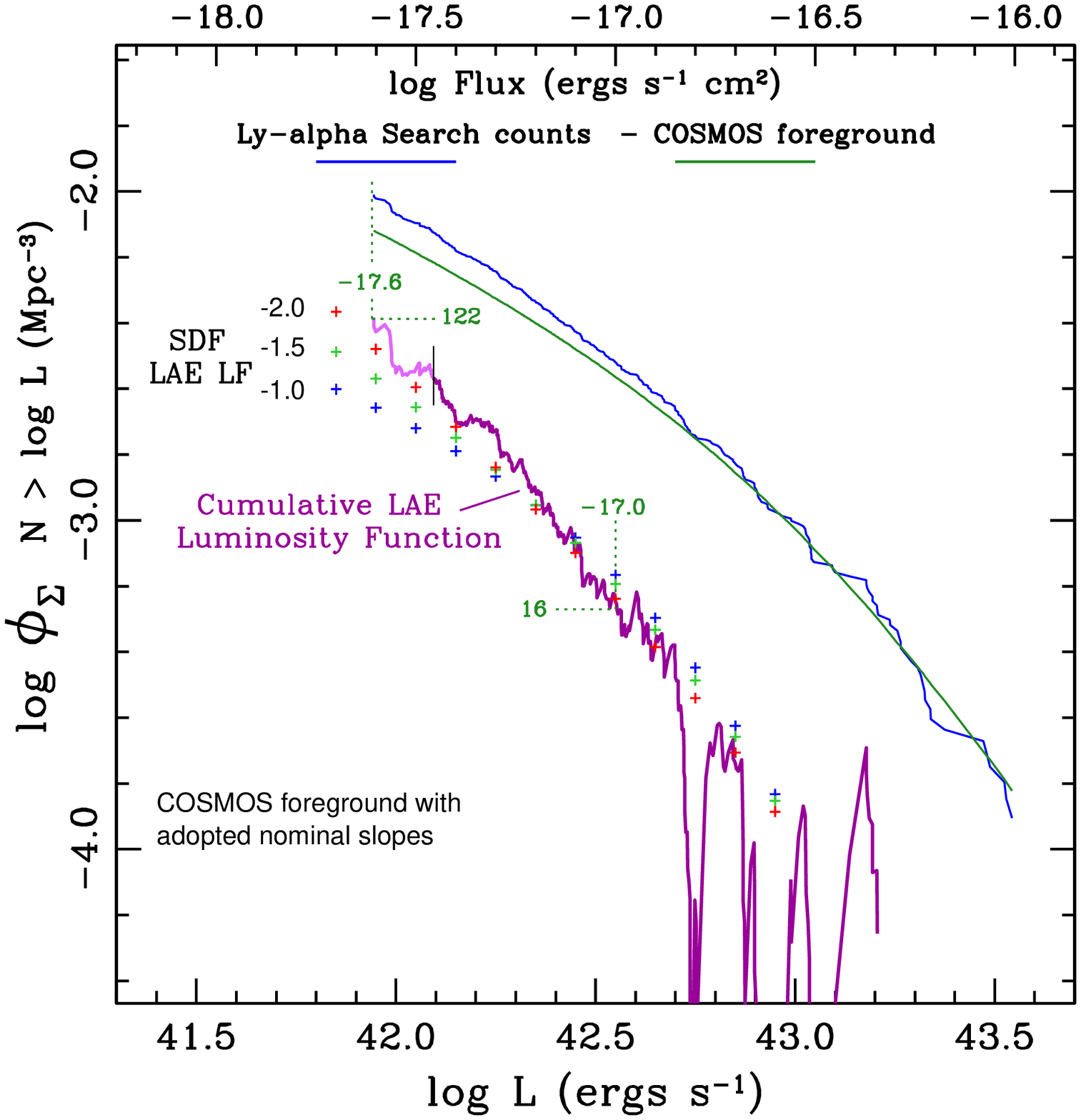,angle=0.,width=3.5in}}

\noindent{\scriptsize
\addtolength{\baselineskip}{-3pt}
 
\hspace*{0.3cm}Fig.~10.\ Cumulative LF for \Lya\ emitters (purple line) derived from subtraction from the search `ALL' counts 
(blue line) of Schechter function fits  to COSMOS foreground counts (green line).    The magenta line is the foreground 
subtracted cumulative LF derived using the nominal Schechter LF slopes of $\alpha =$ -1.30, -1.60, and -1.60 (\OII, \OIII, \& \Ha), 
as described in the text.  As in the previous figures, the data are cumulative rather than binned: this accounts for the wild 
excursions at the bright end of the LF --- points where the predicted foreground exceeds the few counts of our comparatively
small-volume survey. The point of 50\% completeness is marked with a short vertical line and a lighter purple to our survey limit.  
Also shown are SDF LF function extrapolations for the LAEs emitters found in narrow-band searches by Shimasaku \etal\ (2006).  
Our derived LF for LAEs rises in accordance with the \emph{predicted} SDF model for a differential faint-end slope of $\alpha = -2.0$.  
The cumulative counts in our survey area increase from $n \approx 16$ for log $L$ (ergs s$^{-1}$) $> 42.55$ to $n \approx 122$ at 
the limit of our survey, log $L$ (ergs s$^{-1}$) $ = 41.95.$

\vspace*{0.2cm}
\addtolength{\baselineskip}{3pt}
}

We find that normalizing the foreground to the total $n = 640$ sources in the full 1/2\deg\ field of \emph{IMACS} results in a 
cumulative LF with 16 LAEs down to $F = 10^{-17}$\flux, about 6\% higher than the number found by integrating the 
LAE LF of the larger SDF survey (Shimasaku \etal 2006), scaled to the $\sim5$ times smaller volume of our survey of 
30,366 Mpc$^{3}$.  However, this excellent agreement must be in largely fortuitous since the SDF LF includes modeling 
for incompleteness and survey boundaries --- not attempted here --- that make differences at the 10-20\% level.  
Our normalization has additional uncertainty of at least 10\% due to its sensitivity to the amount of subtracted foreground, 
which we do not know better than a few percent due to Poisson fluctuations, and also because we do not as yet have 
independent measurements of the foreground in the 15h search field.  While the agreement in the space density
of LAEs brighter than $F = 10^{-17}$\flux\ gives us confidence in our foreground-subtraction analysis, we adopt the SDF 
normalization and conclude that the agreement of our LAE density with that of SDF is if anything better than expected 
and thus provides no new information on the LAE volume density.

Comparing the slope of putative LAE sources fainter than $F = 10^{-17}$\flux\ with the SDF \emph{predicted} LF over this range, 
our measured LF follows the steepest of the SDF models with slope $\alpha = -2.0$, as Figure 10 shows.  Using this nominal 
model of the foregrounds, the cumulative number of LAEs  rises from 16 at $F = 10^{-17.0}$\flux\ to $n \approx 122$ at $F = 10^{-17.6}$ 
($2.5\, \10-18$)\flux.  The steep rise in LAEs found by statistically subtracting the foreground contamination produces a
result that is not greatly different from the SEL sample (see Figure 9), which means that a substantial fraction of the 
single-emission-line-only sources (compared to a much smaller fraction of \emph{all} emission-line sources) are predicted 
to be LAEs.  The steep rise attributed to LAEs is also consistent with the result of the cross-correlation analysis in \S5.2.  
Although only circumstantial as opposed to the direct evidence of high-resolution spectroscopy, the agreement in result 
for the two fully-independent analyses of (1) LAE LF by background-subtraction, and (2) angular correlation function of 
foreground and candidate LAEs, provides some confidence that approximately half of the faint SEL sources we have 
found are LAEs.  This is the main conclusion of this paper.

The critical reader will remember that the result of the cross-correlation analysis of \S5.2 implied that at  most about half of 
the faint LAE candidates are members of the foreground population.  The foreground subtraction analyses suggests
that at $F = 10^{-17}$\flux\ only about 15\% of all emission-line sources are predicted to be LAEs, while this fraction
rises to $\sim50\%$ at the limit of our survey, $F = 10^{-17.6}$ ($2.5\, \10-18$)\flux.  Thus, the two independent approaches 
yield the same approximate result, though there is still considerable latitude in both estimates.

\subsection{Placing limits on the LAE luminosity function}

The uncertainty to be attached to the LF shown in Figure 10 is not governed by the counts of sources.  Much larger
changes in the derived LAE LF come from the uncertainty in the level of foreground contamination.  First, there is 
the statistical uncertainty in fitting the COSMOS foregrounds counts with Schechter functions.   As described in 
\S5.2, there are a range of slopes (with corresponding values of $L^*$) that can fit the data; in \S5.4 we used 
only the `best fit' values.   Flatter slopes for the foreground LFs will increase the amplitude and steepness 
of the LAE LF, while steeper slopes for the foreground will do the opposite.  The foreground with the most leverage 
is \OII, although we steepen somewhat the slopes of \OIII\ and \Ha\ as well.  The turquoise line in Figure 11 shows 
the result of adopting the $-1\sigma$ fit (see Figure 8) of the Schechter LF for \OII\, $\alpha = -1.05$: even if the \OIII\ 
and \Ha\ slopes are left at their steeper `nominal' values, the flatter \OII\ LF produces an even stronger climb in the 
LAE LF, $\alpha_{LAE} \approx -2.5$, compared to the nominal result of Figure 10 (the magenta line in Figure 11).  
Conversely, a steeper \OII\ slope $\alpha = -1.50$ ($+1\sigma$) with \OIII\ and \Ha\ slopes of $\alpha = -1.65$ 
produces a flatter LAE LF (gold line), but it is still prominent  with $\alpha_{LAE} \approx -1.3$.  Therefore, a rising 
LAE LF --- while quantitatively sensitive to the amount of foreground contamination --- holds over a substantial and 
reasonable range of foreground contamination.

In addition to these statistical errors to the fits in the COSMOS foreground counts, we must consider systematic errors in determining
the foreground that can influence the LAE LF derived by this technique.  We have in fact used only the foregrounds from the
COSMOS survey, but we have applied this foreground to both the 10h (COSMOS) and 15h (LCIRS) fields for which we have
produced counts of emission-line sources with and without continuum.  Cosmic variance is expected and this could have a
substantial effect on our LAE LF.  For example, Ly \etal\ (2007) fit Schechter functions to emission-line source counts uncorrected 
for extinction in the Subaru Deep Field.  Using photometric selection techniques different from the COSMOS study we use
here, Ly \etal\ obtain an almost identical normalization of counts at $F = 10^{-17}$\flux\ compared to COSMOS, and ratios of Ha, OIII, 
and OII of 13\%, 23\%, and 64\%, compared to 13\%, 32\%, 55\% for the COSMOS foregrounds --- an inconsequential difference. 
However, despite very similar  $L^*$ values for these three LFs, the much flatter slopes of the luminosity functions derived by Ly \etal\, 
$\alpha = -1.15$ for \OII and -1.22 for \OIII, result in a 15\% lower foreground at the limit of our study, $F = 10^{-17.6}$\flux, and 
predict an even larger population of LAEs than our nominal result.  Accordingly, there are also considerably higher counts for the 
bright end of the LF compared to COSMOS, which leads to a considerable oversubtraction from all our own MNS counts of 
bright sources.

The Ly \etal\ example again points to faint-end slopes as the key issue.  The bright end of each foreground LF, in contrast, 
is sensitive to the normalization and exponential cutoff of the Schechter LF and not well constrained by the comparatively small 
volume of our survey.  For this reason, although we have not needed to renormalize the COSMOS-derived foregrounds before 
subtracting from our MNS counts, we would have considered it acceptable to do so and to concentrate our attention on the faint end 
slopes, where the case for a steeply-rising LAE LF is made or broken.  Because they are well below the $L^*$ values of these LFs, our 
extrapolations of LF fits to the measured COSMOS counts are essentially power-laws: our result does not depend on the {\it shape} 
of the LF, either theoretically or as measured.

Comparing our own MNS counts in the two fields, we find evidence for both variation and consistency.  Including the 
incompleteness correction in both, our total counts of all emission-line sources are very close at 276 in the 10h field 
compared to 258 in the 15h field.  However, the bright ends of the LF are strikingly different: 19 of the 20 sources brighter 
than $F > 5 \times 10^{-17}$\flux\ are from the the 10h field.  Approximately two-thirds of these are identified to be \Ha\ 
using our short spectra (as is predicted from the COSMOS LFs shown in Figure 9), however, it appears that large-scale 
structure has caused \Ha\ to be over- and under-represented in the 10h and 15 h fields.\footnote{From these 
semi-reliable line identifications from the MNS survey data, for which [N II] $\lambda$6548, $\lambda$6583 and 
[OIII] $\lambda$4959 are sometimes apparent, we determined ratios of 53\%, 31\%, and 16\% for \OII, \OIII, and \Ha, 
in good agreement with the proportions determined in COSMOS.}  Basically, then, the faint end (power-law) slopes of 
the LF fits to the foreground populations dominate the uncertainty in our result of a steeply rising LAE LF, whether this 
comes from statistical errors in fitting the COSMOS counts or in uncertainties that arise from cosmic variance that 
projects different proportions of the three main foregrounds into our two fields.  

Focusing on faint end slopes, then, we find that a complete loss of the LAE signal for our survey requires
pushing the slopes of all three foregrounds to $\alpha = -1.70$.  For the \OII\ LF this is a $+2\sigma$ excursion from 
the best fit of the COSMOS data.  Such steep slopes for the foreground populations can account for 
all of the observed SEL plus Em+C sources.  As seen in Figure 11, the resulting  cumulative LF (the grey line) is flat , which 
means that \emph{no} LAEs are added fainter than $F = 10^{-17}$\flux.  We find this possibility unlikely for three reasons: (1) It is 
inconsistent with the LAE LF found in the SDF, which rises far faster up to $F = 10^{-17}$\flux\ and shows no sign of 
turning over at that point. (2)  It is inconsistent with the lack of correlation between these faint sources and the COSMOS
foreground (\S5.2). (3) It requires ``fine tuning'' the slopes of the foreground population --- a wide range of slopes, 
$-0.85 < \alpha < 1.50$ for both \OII\ and \OIII\ (encompassing the values found by all other studies) lead to a rising LAE LF, 
but only a narrow range $1.60 < \alpha < 1.70$ leads away from this conclusion.  For these reasons we consider a
rising LAE LF to be the most probable interpretation of our MNS survey data, and the one we favor. 


\vspace*{0.1cm}
\hbox{~}

\centerline{\psfig{file=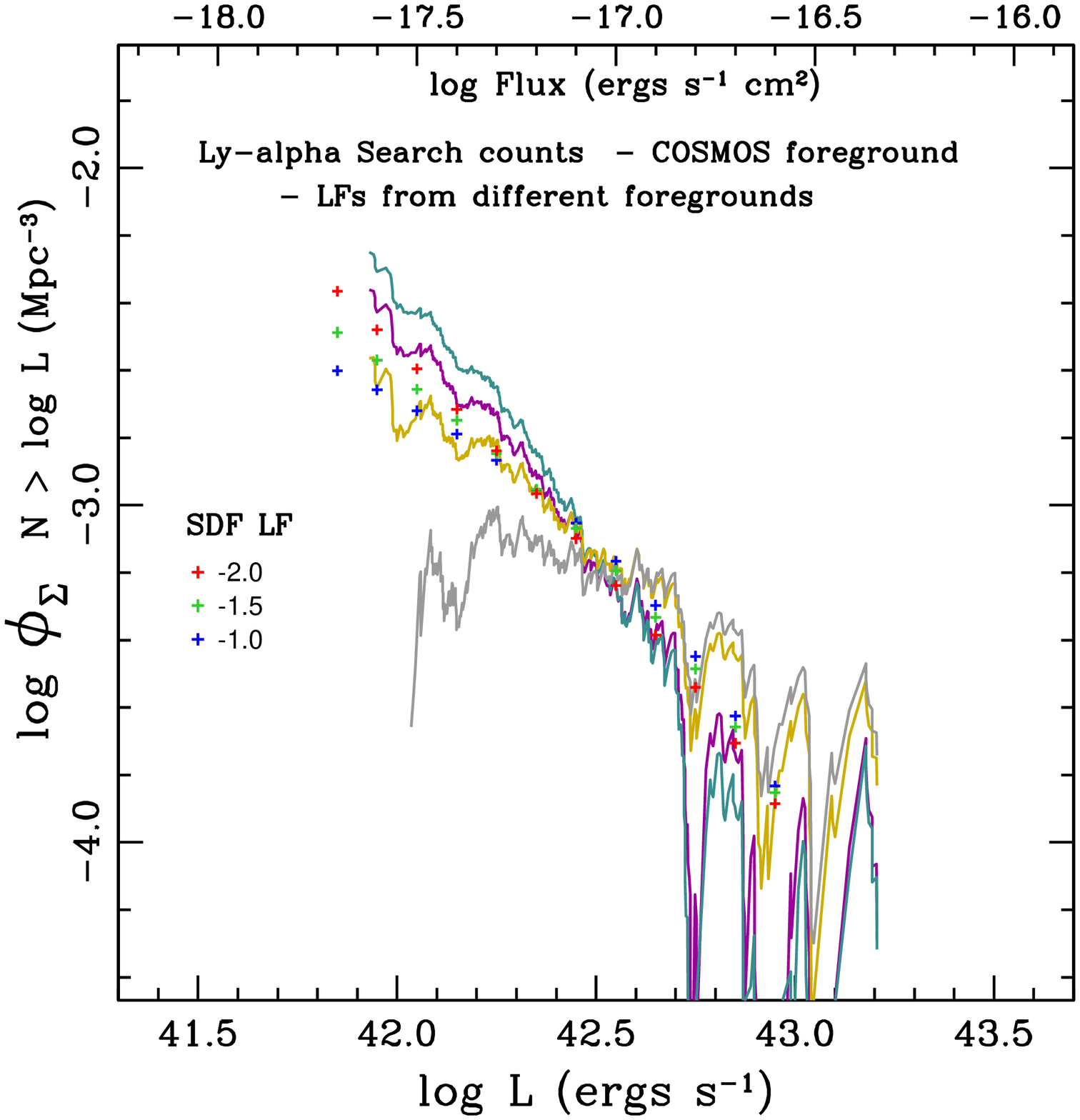,angle=0.,width=3.3in}}

\noindent{\scriptsize
\addtolength{\baselineskip}{-3pt}
 
\hspace*{0.3cm}Fig.~11.\ Same as Figure 10, but for shallower and steeper Schechter-function slopes for the 
foreground populations.  The magenta line is the LF derived for the nominal slopes shown in Figure 10, 
which approximately matches the extrapolated Subaru Deep Field LAE counts with a slope $\alpha_{LAE} \approx -2.0$. 
The LF above (turquoise) has the nominal \OIII\ and \Ha\ slopes but the $-1\sigma$ shallower \OII\ slope of -1.05 (see 
Figure 8), which leads to a steeper LAE LF with $\alpha_{LAE} \approx -2.5$. Steeper slopes of the nominal foreground 
fits, - 1.50, -1.65, and -1.65 (\OII, \OIII, \Ha -- see text) produce the shallower LAE LF (gold line) with $\alpha_{LAE} 
\approx -1.3$.  Within this range of foreground fits the basic result of a rising LAE LF is maintained.  However, steepening 
the foregound poulation slopes further, to -1.70, -1.70, and -1.70 makes the rising LAE LF disappear (grey line), in fact, 
there are no added LAEs fainter than $F = 10^{-17}$\flux.   As explained in the text, there are reasons  to be dubious 
of this model, including its inconsistency with the brighter LAE LF of SDF, the angular correlation results of \S5.2, 
and the ``fine tuning'' it requires. 

\vspace*{0.2cm}
\addtolength{\baselineskip}{3pt}
}
To summarize, we have used the COSMOS foreground counts to subtract from our full ensemble of emission-line 
sources and found a steeply rising LAE LF that is significant at the $\sim2\sigma$ level, where this is set 
by the statistical uncertainty in the foreground contamination rather than by the number of faint emission-line 
sources we have observed.  Systematic errors will play a significant role only if they effect the faint end slopes of 
the LFs for foreground sources.  Within broad limits of the faint-end slopes of the foreground LFs, the result of
a rising LAE LF is secure.  Nevertheless, this exercise has confirmed that measuring the LAE LF faint-slope 
to $\pm$10\% is is probably not possible with this method alone, due to the sensitivity of a rising LAE LF to 
the foreground subtraction.  This highlights again the importance of spectroscopic confirmation for these 
faint candidate LAEs.  While difficult, this can be done with existing facilities and is ongoing.

\section{Discussion}

Early galaxies are central to the hierarchical assembly of present-day galaxies, the dispersal
of heavy elements, and the reionization of intergalactic hydrogen. The number density of low mass
galaxies is arguably the most important unknown in each of these chapters of cosmic history.

The 2008 MNS survey we present in this paper is unprecedented in its combination of depth and 
volume. The faintest $z\approx5.7$ \Lya\ emission lines found in the survey have a line luminosity
of $L(\Lya) = 7.0 \times 10^{41}$~ergs~s$^{-1}$, which is nearly 5 times deeper than narrowband-imaging
surveys for LAEs.  The only known sources of comparable luminosity at $z \sim 5.7$ have either been found 
in the UDF  and UDF Parallels (e.g.,  Bouwens \et 2006; Stanway et al. 2007), behind foreground clusters (Ellis et al. 2001;
Santos \& Ellis 2004), or in the 3.3 square-arcminute `Ultra-deep' serendipitous VIMOS Deep Survey (Cassata \etal\ 
2011).   Our MNS survey covers a far greater volume than these studies.  The measured FWHM filter bandpass of 
($8115-8249$\ang) of our MNS survey samples \Lya\ emission over a redshift interval of $\Delta z =  0.11$ at $z = 5.75$, 
producing an effective survey volume of $1.52\times10^4$  Mpc$^{3}$ per mask --- two-orders-of-magnitude 
larger than the HUDF for an equivalent redshift slice.  Moreover, as a result of the layout of our slit mask, our  survey 
sparsely samples an area of sky 10  times larger than the solid angle subtended by the slits. For the two-field survey, then, 
our total survey volume of  $3.04\times10^4$ Mpc$^{3}$ is actually drawn from a $3\times 10^5$~Mpc$^{3}$ volume, 
substantially reducing the effect of cosmic variance.

The area of the slits used to probe gravitational lens caustics in clusters of galaxies is also tiny 
compared to our MNS mask, and magnification by the lens further reduces the survey volume. 
For example,  Santos \& Ellis (2004) present survey volumes up  to $10^4$~Mpc$^{3}$ over the 
a broad redshift range corresponding to $\Delta z \approx 2$. In contrast to this pencil-beam 
footprint, our survey volume is more nearly cubical, making our MNS survey much better suited
to constraining the faint-end slope of the LAE luminosity function. 


\subsection{The Connection of Low Luminosity LAEs to Galaxies like the Milky Way}

While mindful of the critical importance of spectroscopic confirmation of our LAE candidates, we adopt for purpose 
of the following discussion the nominal LAE LF shown in Figure 10 based on the most probable foreground subtraction.
Our MNS survey finds a surface density $N \gs 1$ LAE candidate per square arcminute, for log $L (\lya)$ (ergs s$^{-1}$) 
$> 41.95$.  With a redshift interval of   $\Delta z =  0.11$, this amounts to a space density of approximately 
$4\,\times10^{-3}$~Mpc$^{-3}$, comparable to that of today's \emph{L*} galaxies. For example, Loveday 
\etal\ (1992) and Marzke \etal\ (1994) find Schechter-function values $\phi^* \approx 5-10\,\times10^{-3}$ Mpc$^{-3}$ 
for the local galaxy population. Integrating the LF with the slope $\alpha=-1.0$ found by these studies, the space 
density of galaxies brighter than about $L^{*}/3$ in the local universe matches the comoving space 
density of our candidate LAEs. Stated another way, we have found three or four sites of active  star formation 
per future $L^*$ galaxy.


%
%
%
%
%

Today's \emph{L*} galaxies have typical halo masses of $M = 1-2\,\times 10^{11}$ \Msun. 
From analyses of merger trees in $\Lambda$CDM simulations of the growth of structure, Stewart \etal\ 
(2008) find that the halo mass of such objects at $z \sim 6$ is an order-of-magnitude smaller, 
$M = 1-2\,\times 10^{10}$ \Msun, and that these will grow into today's typical galaxies principally through 
accretion of smaller halos of $\sim$10\% of their mass. Interestingly, a halo mass of $\sim10^{10}$ \Msun\
can also be inferred for LAEs at the density found by our survey by comparing to the LAE study by Ouchi \etal\ 
(2010), who derive a halo mass of $M = 10^{10-11}$ \Msun\ from the clustering strength of a large sampe of brighter 
LAEs, \emph{$L> L^*$}.  Although our sample is too small, and not well constructed to measure correlation
statistics, it is reasonable to assign halo masses an order-of-magnitude smaller to the much less luminous LAEs 
of our study, $\sim10^{9-10}$ \Msun.  This in turn suggests a stellar mass of $\sim10^{8-9}$ \Msun.



The fraction of all galaxies that are detectable as LAEs at this redshift remains uncertain,
so sampling issues are another important consideration in this comparison. Our survey 
ha only recovered  galaxies with vigorous star formation and exhibiting
\Lya\ emission. These LAEs could be short-lived starbursts,  so we might only be detecting 
a fraction of such galaxies at this mass. Ouchi \etal\ estimate a  $\sim$1\% duty cycle,
but Stark \etal\  (2011) argue for a value much closer to unity, as perhaps does
our own finding of a steep faint-end slope of $\alpha\sim-2.0$ for the LAE LF, if this
is a manifestation of \Lya\ luminosity following halo mass.  Whether large or small, the 
correction for duty cycle will surely increase the luminosity density our MNS survey implies.

In addition,  we may be finding only the fraction of active objects to which our view is not obscured 
by dust.   Although there is some reason to expect that this is becoming less of a factor for 
higher redshift, lower-luminosity systems (Bouwens \etal\ 2009), here again we can be sure of the 
sign of the effect, and in this case the steepness of faint-end slope does not limit that this effect 
to being a small one.  In summary, we can be fairly certain that our survey identifies only a fraction 
of the galaxies of this dynamic period of star formation and assembly, implying
a higher space density of galaxies for a population that --- according to our result --- already 
outnumber future $L^*$ galaxies by a factor of $3-4$.

By number density arguments, then, we can be confident that faint LAEs are not components of
the most massive objects at $z \sim 6$, but more likely progenitors of today's \emph{L*} galaxies.  
However, it is less clear whether these LAEs are in fact part of the main structure of such a galaxy 
(e.g., the early halo or bulge) or instead some accreted satellite or lower mass dwarf system that is destined 
to become part of the \emph{L*} galaxy.   Significant advances in understanding what role these systems play 
are beyond present observational capabilities.  Eventually, however, with measurements of stellar and dynamical 
masses, and large enough samples for an accurate correlation function statistics, the nature of this basic, 
populous component at a crucial epoch of galaxy growth will become clear.


\subsection{The Properties of Galaxy Building Blocks}

While \Lya\ luminosity is admittedly a poor indicator of star formation rate, due to radiative transfer 
effects in galaxies, we note that the standard case~B conversion gives star formation rates as low as 
SFR $= 0.84 f_{Lya}^{-1} (F($\flux)$/2.5 \times 10^{-18})$ \Msun\ $yr^{-1}$ for the faintest objects in
 our new sample of LAEs.  If star formation lasts for at least a dynamical timescale, $\tau \gs\ 5.0 \times 10^7 yr$,
 and half the \Lya\ photons escape, $f_{\Lya} \approx 0.5$, then these galaxies must have
built up  stellar masses of at least $10^8$ \Msun, consistent with what we derived above based on 
estimates of halo masses of $\sim10^{9-10}$ \Msun.

It is interesting to compare this mass scale to some of the earliest building blocks of present-day galaxies. 
The surviving Milky Way satellites and globular clusters (GCs) may be the oldest intact stellar systems that formed 
near the end of the reionization era.  The characteristic stellar mass scale of globular clusters today is 
$1.4 \times 10^5$ \Msun\ (Harris 1991).  Even if the initial masses of GCs are, as estimated, at least 
8-25 times larger (Schaerer \& Charbonnel 2011), it still follows that the faint LAEs probably contain more 
stellar mass than early globular clusters. By this argument, the faint LAEs discussed hear probably represent 
either the main body of a proto-Milky-Way galaxy or its satellite galaxies.

We can also consider the fate of the heavy elements synthesized by galaxies near the end of
the reionization era. From the distribution of intervening metal-line-systems at redshift $z\sim3$,
we know that the metals ejected by galaxies present a high clustering bias. While the LBGs studied by Adelberger \etal\ 
(2005) are similarly clustered and have been shown to harbor strong outflows (Steidel \etal\ 2010), 
their stellar populations are not generally old enough to have spread metals over the large enriched 
regions recently identified using tomographic studies of intervening metal-line systems (Martin 
\etal\ 2010).  These authors argue that lower mass galaxies at higher redshift dominate the 
dispersal of heavy elements and show that winds from redshift $z\sim6$ galaxies with halo
masses of $0.5-2.0 \times 10^{10}$ \Msun\ would have left very large bubbles of metals around 
LBGs at $z \sim 3$.  Therefore, by parameters of space density and mass, our faint LAE sample 
appears to be a good match to this hypothesized galaxy population.  We therefore expect strong
gaseous outflows from these objects, a prediction that can be tested with additional spectroscopy. 
Since these enriched bubbles may persist long after the star formation in these galaxies, their redshift 
density could eventually constrain the duty-cycle of low-luminosity LAEs.

\subsection{Faint LAEs, the Lyman-continuum, and Reionization at $z \sim 6$}


The production of Lyman-continuum photons by galaxies at $z\sim6$ has been of particular interest 
since the discovery that intergalactic hydrogen is almost entirely ionized by this time (Becker \etal\ 
2001; Fan \etal\ 2002).  \emph{Wilkinson Microwave Anisotropy Probe} measurements point to a 
significant ionization fraction by around $z\sim11$ (Spergel \etal\ 2007; Page \etal\ 2007), but the 
recombination time of intergalactic protons is short compared to the time between $6 < z < 11$.  An 
ionized IGM at $z=6$, therefore, requires a critical density of star formation or another, as yet unidentified, 
source of Lyman-continuum radiation.

For many years it has been recognized that low luminosity galaxies at $z \gs 6$ played an important role
in the reionization of the universe (Lehnert \& Bremer, 2003; Stanway \etal\ 2003; Bouwens \etal\ 2007).
These studies found many galaxies at $5 < z < 7$ that are fainter than the knee in the luminosity, but the 
implied star formation density of the detected galaxies falls well below the critical level required to
maintain reionization.  Thus, even more sensitive surveys are required to identify this lower-luminosity 
population of galaxies that maintains reionization.  Our MNS survey has taken a significant step in finding 
faint \Lya\ emission from galaxies  whose continuum radiation could be undetectable with any available facility.

With this as background, we revisit the question explored in Martin08 of the possible contribution of faint LAEs 
to the reionization budget of the IGM at $z\sim6$.  As discussed in Martin08, the critical star-formation-rate-density at 
$z \sim 6$ required to keep the intergalactic hydrogen ionized depends  on the physical properties of galaxies and the IGM. 
The production rate of Lyman-continuum photons, the fraction of Lyman-continuum photons that escape from galaxies 
($f_{LyC}$), and the clumpiness of the IGM (\emph{C}) may all evolve significantly over the first few billion years  of galaxy 
formation. Martin08 grouped these dependencies into a single parameter 
$\zeta \equiv C_{6} (1 - 0.1 f_{LyC,0.1}) f_{Ly\alpha,0.5} / f_{LyC,0.1}$; $\zeta\sim1$ corresponds to the current 
best-guesses for these parameters. 
Values of $\zeta < 0.2$, or $\zeta >  5.0$, would require significant modifications to our current understanding of 
these physical properties.\footnote{Here, we have parameterized $C_6 = C/6$, and $f_{LyC, 0.1} = f_{LyC} / 0.1$.  
Likewise, the escape fraction of \lya\ photons is given as $f_{Ly\alpha,0.5} = f_{Ly\alpha}/0.5$.  The  `best-guess' values 
of these  renormalizations are discussed in more detail in Martin08.} In addition, the intergalactic gas 
temperature of $10^4$ K we assume is also conservative. Higher temperatures would lower the recombination rate 
and thereby reduce the number of Lyman continuum photons required for reionization.


Martin08 used the LAE luminosity function parameters --- \emph{$L^*$}, $\phi^*$, and $\alpha$ --- to estimate the number 
of ionizing photons.  The production of Lyman-continuum radiation by stars is quite uncertain for any individual LAE.  For 
the population as a whole, however, under the standard Case~B recombination assumption, a galactic nebula produces 
two \Lya\ photons for every three Lyman-continuum photons absorbed by the galaxy; we assume that, on average, one 
of the two \Lya\ photons escapes from the galaxy, i.e., $f_{\Lya} = 0.5.$ We adopt the `best fit' faint-end slope of $\alpha = -2.0$ 
and the $\pm1\sigma$ bounds, $-1.3>\alpha>-2.5$, that come from varying the foreground contamination (see \S5.3) in our 
new MNS survey.  Shimasaku \etal\ (2006) give fits to \emph{$L^*$} and $\phi^*$ for slopes $\alpha=-1.0, -1.5, -2.0$.  We 
interpolated, and extrapolated, respectively, to find appropriate [$L^*$, $\Phi^*$] values for the  $\alpha=-1.3$ and $\alpha=-2.5$ limits.  
For the slopes of $\alpha=-1.3, -2.0, -2.5$ the adopted values are determined as log $L^*$ (ergs s$^{-1}$) $= 42.81, 
43.20, 43.63$ and log $\phi$ (Mpc$^{-3}$) = -3.05, -3.80, -4.71.  The three curves in Figure 12 show the integration of flux 
density to decreasing luminosity limits and compares this to the value $\zeta$ described above that represents the 
required luminosity density to keep the IGM ionized.

A firmer constraint on the LAE LF allows us to better estimate the fraction of the critical ionizing flux
that faint LAEs contribute to maintaining ionization at $z\sim5.7$.  Our new data indicate a steep 
faint-end slope  $\alpha\sim-2$ (see Figure 10), so for purposes of our discussion here, we revisit only 
the luminosity-density calculation shown in Figure 13 of Martin08 for $\alpha=-2.0$.  That diagram
showed a strong covariance of $L^*$ and $\Phi^*$ that produced a banana-shaped region of 
values that were compatible with the few data of that analysis.  However, if we use our new MNS results
to constrain the faint-end slope and adopt the [$L^*$, $\Phi^*$] values obtained by Shimasaku \etal\ (2006) 
for Schechter-function fits for the SDF data, the analysis is more straightforward.  We recognize that
both the faint-end slope from this work and the $L^*$ and  $\Phi^*$ values from the SDF survey continue 
to have significant uncertainties, so our analysis here is only for the best values available.

In Figure 12 we plot the integrated luminosity density --- $\mathcal{L}_{\rho}$ --- for the LFs with faint-end slopes
$\alpha = -1.3, -2.0, -2.5$ shown in Figure 11.  Summing the \Lya\ luminosity down to log $L$ (ergs s$^{-1}$) $= 42.55$ 
falls short by an order-of-magnitude of the $\zeta=1$ case.  Our new data are relatively complete down to  log 
$L$ (ergs s$^{-1}$) $= 41.95$. With $\alpha = -2.0$ and the Shimasaku et al. normalization, our new population of 
LAEs comes up  a factor of five short for maintaining the ionization of the IGM at $z=5.7$.  However, looking at this 
result another way, if the escape fraction of Lyman-continuum photons from these galaxies were 50\% ($\zeta = 0.2$) 
instead of the assumed value of 10\%, then the population that we detected can fully ionize the IGM. 

For purposes of further discussion we assume that the LAE LF slopes we derive hold to much fainter luminosities.  For 
example, our best-fit value of $\alpha = -2.0$  for the LAE LF needs to hold to a luminosity 28 times fainter --- to log 
$L$ (ergs s$^{-1}$) $= 40.5$ --- to reach critical flux density for an escape fraction of 20\%, $\zeta=0.5$.  With the 
steeper faint-end slope of $\alpha = -2.5$ --- consistent with our survey result at the $1\sigma$ level --- our counts 
are already sufficiently deep to reach critical flux density $\zeta=0.5$, and even $\zeta=1.0$ requires only a factor 
of $\sim$7 extension to fainter LAEs, to log $L$ (ergs s$^{-1}$) $= 41.1$.


\vspace*{0.1cm}
\hbox{~}

\centerline{\psfig{file=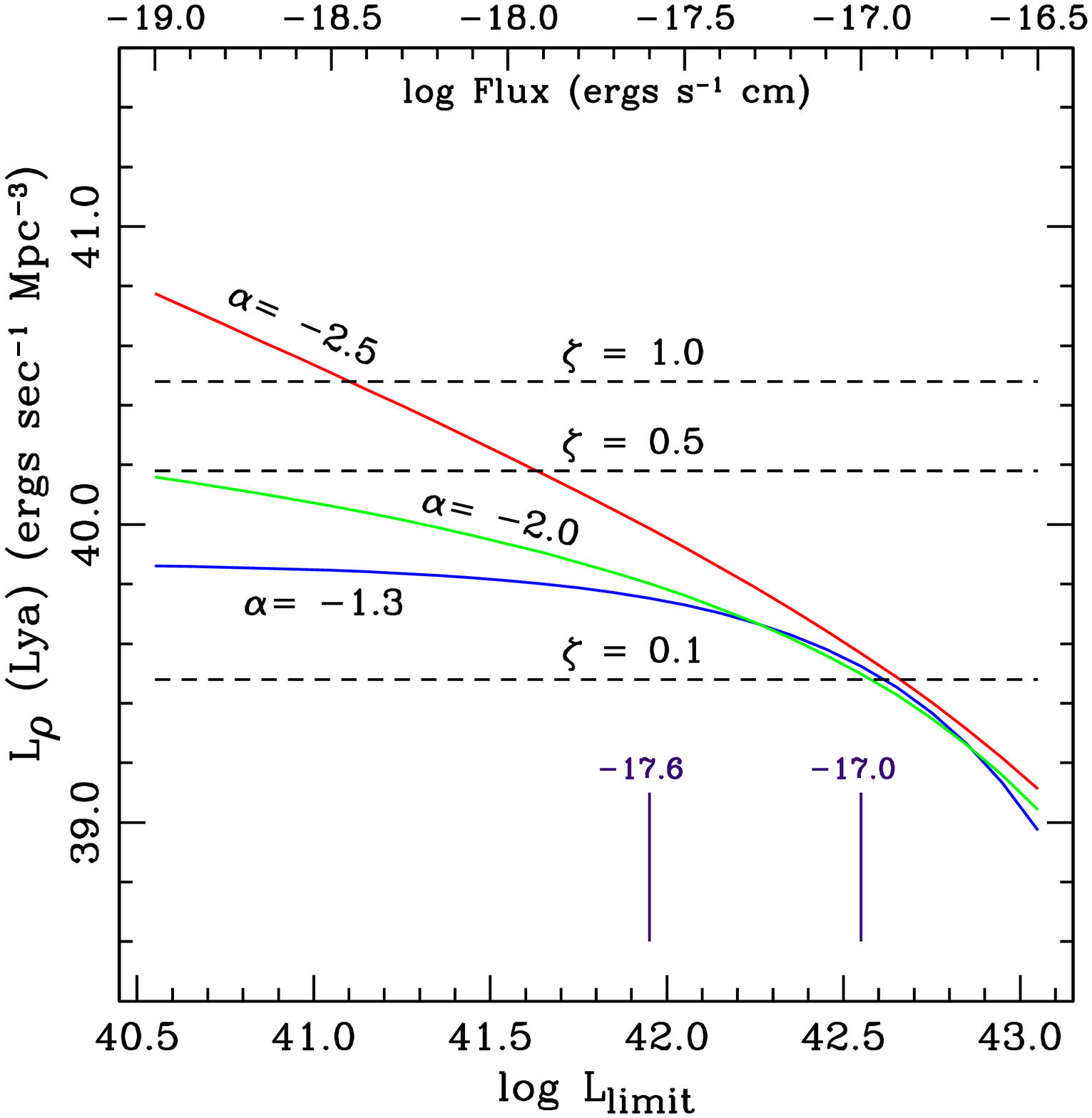,angle=0.,width=2.8in}}

\noindent{\scriptsize
\addtolength{\baselineskip}{-3pt}
 
\hspace*{0.3cm}Fig.~12.\ Level of luminosity density required for maintaining reionization at $z\sim5.7$ with a
population of faint LAEs.  The green, blue, and red curves show the integrated luminosity density for the three
values of faint-end slope $\alpha$, -1.3, -2.0, and -2.5, respectively, that characterize the `best fit' and $\pm1\sigma$
bounds of our LAE LF result.  For slopes $\alpha<-2.0$ there is substantial progress toward reaching critical flux 
density, $\zeta=1$, from our prior limit (Martin08) of the LAE LF, log $L$ (ergs s$^{-1}$) $\sim 42.55$, to the lower 
luminosity limit of the MNS survey.  By log $L$ (ergs s$^{-1}$)) $\sim 41.95$ about 20\% if the critical flux density 
$\zeta=1.0$ has been reached with the $\alpha=-2.0$ slope, or about 40\%, if the Lyman-continuum escape fraction 
is as high as 20\%, or \emph{all}, for an escape fraction of about 50\% --- not impossible for these low-luminosity, 
high-z LAEs.  Extrapolation to fainter limits, and/or a steeper slope of $\alpha=-2.5$, further increase the likelihood 
that LAEs alone can provide the critical flux density to complete reionization at $z\sim6$. 
 
\vspace*{0.2cm}
\addtolength{\baselineskip}{3pt}
}

If the rising LF for faint LAEs we have derived is correct, an important step has been taken towards finding the galaxies 
that could maintain the ionization of the IGM at $z=5.7$. At  $z \sim 6$, however, direct measurement of the 
Lyman-continuum escape fraction is precluded due to the attenuation of the Lyman-continuum by intergalactic hydrogen. 
Proof that the faint LAEs galaxies can maintain the ionization of the IGM will rest on indirect arguments about their 
Lyman-continuum escape fraction. One approach is to extrapolate from direct measurements at lower redshift. Another 
approach is to model the nebular emission.

In contrast to the normal situation, where the far-UV continuum is dominated by stellar continuum, the nebular 
continuum is several times brighter than the stellar continuum for extremely young star clusters (see Figure 5 of 
Schaerer 2002). Even for less extreme conditions, the nebular continuum makes a significant contribution to the 
far-UV luminosity. The hydrogen two-photon continuum makes the galaxy appear redder than its intrinsic stellar 
SED. Figure 3 of Bouwens et al. (2010) illustrates this effect as a function of stellar age and suggests that less 
extreme nebular emission still impacts the continuum slope. Most significantly, nebular emission prevents the 
continuum slope from getting bluer than about $\beta \sim -2.5$.\footnote{The far-ultraviolet continuum slope 
is defined by $f_{\lambda} \propto \lambda^{\beta}$ or $f_{\nu} \propto \nu^{-(\beta + 2)}$ (Meurer \etal\ 1999).}
This limit is relevant because bluer values of $\beta$ have been reported in low luminosity galaxies
at high redshift (Bouwens \etal\ 2009, 2010). The escape of Lyman-continuum photons, by definition, 
reduces the nebular emission and offers one plausible explanation of extremely blue far-UV color. 
In summary, if it turns out that our faint LAEs exhibit very blue far-UV continuum slopes, e.g., $\beta < -2.5$, this 
would be an argument that the escape fraction of Lyman continuum photons in these galaxies is significant.


\section{Conclusion}

We carried out a blind, wide-field emission-line survey, detected line fluxes as low as $2 \times 
10^{-18}$\flux, and discovered what appears to be a steep rise in the number counts of emission-line 
galaxies. Extrapolation of the number counts for well-measured, foreground populations suggests that 
$z=5.7$ LAEs, reaching $L = 7\times 10^{41}$\lum, comprise a substantial fraction of the faint counts 
of the MNS survey sample,  and that the implied faint-end slope of the LAE luminosity  function is close 
to $\alpha = -2.0$. The detected galaxies would then be a major contributor to the sources of 
Lyman-continuum that completed the reionization of the universe at $z\sim6$.  Because of their high 
number density --- several times that of today's $L^*$ galaxies (and even higher, if such factors as 
duty cycle and preferred viewing angles are important), these faint LAEs are likely building 
blocks of today's common galaxies.

It should be possible to estimate the escape fraction of Lyman-continuum photons from measurements 
of the UV-continuum slope of these faint LAEs. Extremely blue values of the continuum slope $\beta$ in 
the rest-frame, far-UV between \Lya\ and roughly 2000\,\AA\ would provide evidence for suppressed 2-photon 
continuum emission and, indirectly, indicate a high escape fractions of Lyman-continuum photons.  Such 
measurements are already possible by using WFC3 on the Hubble Space Telescope to obtain very deep 
infrared photometry of these fields. Measurements of this type with greater depth and quality will 
become routine with the James Webb Space Telescope.

 \section{Acknowledgments}
 
The authors thank the referee for a careful reading of the manuscript and many useful suggestions.
They also acknowledge important help and discussions by colleagues Peter Capak, T.\,J. Cox, Janice Lee, 
and Chun Ly, and comments from Kristian Finlator and Moire Prescott on the presubmission
draft.  Dressler recognizes the critical contribution to this work by the National Science Foundation through
 a 2006 \emph{TSIP} grant, administered by AURA/NOAO, that supported construction of a new mosaic CCD 
camera and made these observations possible. CLM acknowledges support for this project through the David and 
Lucile Packard Foundation. This research has made use of the NASA/ IPAC Infrared Science Archive, which is 
operated by the Jet Propulsion Laboratory, California Institute of Technology, under contract with the National 
Aeronautics and Space Administration.

\end{document}